\documentclass[10pt,journal,compsoc]{IEEEtran}

\usepackage[utf8]{inputenc}

\ifCLASSOPTIONcompsoc
  \usepackage[nocompress]{cite}
\else
  \usepackage{cite}
\fi

\usepackage[hidelinks]{hyperref}
\usepackage{doi}

\ifCLASSINFOpdf
  \usepackage[pdftex]{graphicx}
  \graphicspath{{figures/}}
  \DeclareGraphicsExtensions{.pdf,.png,.jpg,.jpeg}
\else
  \usepackage[dvips]{graphicx}
  \graphicspath{{eps/}}
  \DeclareGraphicsExtensions{.eps}
\fi

\usepackage[caption=false,font=normalsize,labelfont=sf,textfont=sf]{subfig}

\usepackage{amsthm, amsmath, amsfonts, amssymb}
\usepackage{algorithmicx, algpseudocode}

\usepackage{array}
\usepackage{booktabs}
\usepackage{multirow}
\usepackage{float}

\newcommand{\UDSfull}{Union Driver Set}
\newcommand{\UDSabbr}{UDS}
\newcommand{\MinUDSfull}{Minimum Union Driver Set}
\newcommand{\MinUDSabbr}{MinUDS}

\newcommand{\CLAPfull}{Cross-Layer Augmenting Path}
\newcommand{\CLAPabbr}{CLAP}

\newcommand{\algoCLAPS}{\textsc{CLAP-S}}   
\newcommand{\algoCLAPG}{\textsc{CLAP-G}}   
\newcommand{\algoRSU}{\textsc{RSU}}        
\newcommand{\algoILP}{\textsc{ILP-Exact}}  

\newtheorem{definition}{Definition}
\newtheorem{theorem}{Theorem}
\newtheorem{lemma}{Lemma}
\newtheorem{corollary}{Corollary}
\newtheorem{proposition}{Proposition}
\newtheorem{problem}{Problem}
\newtheorem{remark}{Remark}

\title{Optimized Control of Duplex Networks}

\author{Haoyu~Zheng,
        Xizhe~Zhang%
\IEEEcompsocitemizethanks{
  \IEEEcompsocthanksitem Early Intervention Unit, Department of Psychiatry, Nanjing Brain Hospital, Nanjing Medical University, Nanjing 210029, China.
  \IEEEcompsocthanksitem School of Biomedical Engineering and Informatics, Nanjing Medical University, Nanjing 211166, China.
  \IEEEcompsocthanksitem Corresponding author: Xizhe Zhang (zhangxizhe@njmu.edu.cn).
}%
}

\IEEEtitleabstractindextext{
\begin{abstract}
Many real-world complex systems can be modeled as \emph{multiplex} networks, where each layer represents a distinct set of interactions among the same entities. Controlling such systems, defined as steering the system to desired states with external inputs, is crucial in various domains. However, existing network control theory largely focuses on single-layer networks, and applying separate controls to each layer of a multiplex system often yields a redundant set of driver nodes, increasing cost and complexity. To address this gap, we formulate the Universal \MinUDSfull\ (\MinUDSabbr) problem for duplex networks, which seeks the smallest set of driver nodes that simultaneously control both layers. We propose a novel algorithm, \emph{Shortest \CLAPfull\ Search} (\algoCLAPS), that efficiently navigates the combinatorial search space of control configurations. By introducing \emph{\CLAPfull} (\CLAPabbr), \algoCLAPS\ iteratively realigns each layer’s MDS to maximize their overlap. We prove the algorithm’s global optimality and demonstrate its efficiency on both synthetic networks and real-world multiplex systems. The results show that \algoCLAPS\ consistently outperforms baseline approaches by reducing the number of required driver nodes and computational time by an order of magnitude. This work provides a powerful, general-purpose tool for optimizing control strategies in multi-layer networks, enabling more economical interventions in diverse fields.

\textit{Code and data.} Our implementation of \CLAPabbr-S and all baselines, together with scripts to reproduce the figures and tables, is available at \url{https://github.com/njnklab/CLAP-S_Algorithm}.

\end{abstract}
}

\begin{document}

\maketitle
\IEEEdisplaynontitleabstractindextext

\IEEEraisesectionheading{
\section{Introduction}
\label{sec:introduction}
}

The study of complex systems, from the intricate web of molecular interactions within a cell to the vast architecture of the internet, is a central theme in modern science \cite{newman_networks_2018, strogatz_exploring_2001}. Network science provides a powerful mathematical framework for modeling these systems, where entities are represented as nodes and their interactions as edges. A fundamental challenge in this domain is network control: the ability to steer a system's dynamics from any initial state to any desired final state through external inputs applied to a subset of its nodes \cite{ruths_control_2014}. Achieving this goal with minimal intervention is critical for applications ranging from designing therapeutic strategies in systems biology to implementing efficient interventions in social networks \cite{gu_controllability_2015, liu_controllability_2011}.

The modern era of network control began with the structural controllability framework, which determines a system's controllability based solely on its network topology \cite{ching-tai_lin_structural_1974}. A seminal contribution by Liu et al.\cite{liu_controllability_2011} demonstrated that identifying the \textbf{Minimum Driver Set (MDS)} required for structural control is equivalent to solving a maximum matching problem on the network's corresponding bipartite graph. This elegant mapping rendered the problem computationally tractable and spurred a wave of research. Subsequent work in single-layer networks has focused on refining this concept, developing algorithms to find all possible MDS configurations \cite{zhang_efficient_2017}, defining metrics like control capacity to quantify a node's importance across these configurations \cite{jia_control_2013, jia_finding_2022}, and exploring diverse control strategies, such as altering indispensable proteins \cite{zhang_altering_2018}, targeting specific communities \cite{piao_strategy_2015}, adapting to dynamic networks \cite{pan_adaptive_2025}, or modifying the network's edge structure \cite{zhang_altering_2019, zhang_altering_2021}. These theoretical advances have been paralleled by impactful applications, from identifying cancer-keeper genes as therapeutic targets \cite{zhang_cancer-keeper_2023} and guiding the design of brain stimulation therapies \cite{gu_controllability_2015, dimulescu_structural_2021}, to bridging control theory with influence maximization in social networks \cite{sadaf_bridge_2024}.

However, most real-world systems are not isolated monoliths; they are inherently multi-layered \cite{kivela_multilayer_2014, boccaletti_structure_2014}. A cell operates through interconnected layers of gene regulation and protein interactions \cite{yuan_structural_2023}; the brain comprises layers of structural and functional connectivity \cite{gu_controllability_2015}; and modern society is woven from multiplex social and communication networks. Applying control theory to such systems introduces significant new challenges. A naive approach of finding an MDS for each layer independently and then taking their union is conceptually simple but practically inefficient, often leading to a redundant and costly set of driver nodes. The core challenge of multi-layer control lies in designing a unified and efficient strategy under the realistic constraint that a single physical entity is either controlled across all layers or in none \cite{menichetti_control_2016}.

Early work on structural controllability established the matching-based view of single-layer networks \cite{ching-tai_lin_structural_1974, liu_controllability_2011}. Building on this foundation, researchers characterized the multiplicity of minimum driver sets (MDS) via enumeration and sampling, leading to tools such as all-MDS enumeration and control-capacity–style statistics used to assess how often a node appears across valid MDS configurations \cite{zhang_efficient_2017, jia_control_2013}. These single-layer advances inspired diverse design strategies—e.g., modifying edges, constraining control-chain lengths, or adapting to temporal/dynamic settings—to trade off input count, energy, and practicality \cite{zhang_altering_2019, zhang_altering_2021, alizadeh_lcc_2023, pan_adaptive_2025}.

For multilayer systems, several lines of work analyze how layering changes controllability and input placement. A central thread extends the maximum-matching framework to multiplex settings and shows that cross-layer structure can fundamentally reshape driver requirements, sometimes stabilizing configurations that are not available in separate layers \cite{menichetti_control_2016}. Another thread studies two-layer (duplex) networks with different timescales, proving that which layer receives inputs dramatically alters the minimum input count—favoring the faster layer under separation \cite{posfai_multitimescale_2016}. Beyond full-state controllability, recent studies explore target controllability and heterogeneous high-dimensional nodes in multilayered systems, emphasizing how inter-layer couplings can render a system controllable even when an individual layer is not \cite{jiang_multilayer_2022, wang_target_2024, wang_multirelational_2024}. There is also work constraining all drivers to a single designated layer or otherwise tying input locations to one layer, which addresses practical placement limits but differs from our objective of jointly optimizing across layers \cite{zhang_peripheral_2016, li_singlelayer_2025}. Complementary strands investigate energy and spectral alignment across layers or spreading-process control, again highlighting structural interactions across layers rather than layer-wise unions of driver sets \cite{srivastava_underpinnings_2021, bernal_spreading_2020}.

A separate body of literature tackles multilayer controllability via alternative graph-theoretic surrogates, most notably minimum dominating sets (MDS-D) and feedback vertex sets (FVS). Dominating-set approaches yield multilayer driver heuristics for undirected or simplified settings \cite{nacher_find_analyse_2019}, whereas FVS-based control targets nonlinear dynamics and attractor steering. Notably, Zheng \textit{et al.} formalized a \emph{minimum union} objective in the nonlinear multilayer setting—minimizing the union of layer-wise FVS driver sets—and proposed greedy procedures for that union minimization \cite{zheng_bmc_2019}. These paradigms optimize different control objectives (domination or cycle breaking) from the matching-based structural controllability we pursue; nevertheless, they motivate our emphasis on minimizing cross-layer redundancy.

Against this landscape, our focus is the \textit{union minimization of matching-based MDS across duplex networks}. Instead of requiring identical drivers across layers \cite{menichetti_control_2016} or restricting inputs to a single layer \cite{li_singlelayer_2025}, we allow each layer to have its own MDS and aim to \emph{reduce the size of their union}. Methodologically, we leverage deterministic, graph-theoretic reconfigurations of layer-wise maximum matchings to increase overlap across layers. Empirically, we compare to random-sampling style baselines that draw multiple maximum matchings per layer and pick the smallest observed union—an approach aligned with prior uses of sampling to probe the space of MDS configurations and node participation frequencies \cite{jia_control_2013}. Our results (Sections \ref{sec:experiments}) show consistent reductions in required drivers and runtime across synthetic and real multiplexes, indicating that targeted cross-layer reconfiguration can substantially outperform undirected sampling strategies.

This work makes three contributions to the study of structural controllability in duplex networks:
\begin{itemize}
  \item \textbf{Graph-theoretic framework for cross-layer controllability.} We formalize cross-layer augmenting paths (\CLAPabbr) that trigger budget-preserving driver exchanges across layers, and show that the absence of any \CLAPabbr\ certifies \emph{global optimality} for the union objective (the \emph{\CLAPabbr-or-Optimal} result).

  \item \textbf{Deterministic algorithm based on shortest \CLAPabbr\ search.} Our procedure \algoCLAPS performs a layer-alternating BFS to find a shortest \CLAPabbr; each successful exchange strictly reduces $\Delta$ by $2$ (hence $|U|$ by $1$) while preserving $(k_1,k_2)$, and halts with a verifiable optimality certificate when no \CLAPabbr\ exists.

  \item \textbf{Extensive empirical evaluation on synthetic and real multiplexes.} On Erd\H{o}s–Rényi, Barabási–Albert, and hybrid duplex networks, as well as diverse real-world systems (biological, neuronal, social, and human relationship networks), \algoCLAPS achieves \emph{exactly the same} optimal union sizes as an exact integer linear programming (ILP) formulation on every instance, yet runs substantially faster in wall-clock time.
\end{itemize}

\section{Preliminaries and Problem Definition}\label{sec:preliminaries}

In this section, we establish the theoretical foundation for our work. We begin by introducing basic graph notations and the principles of structural controllability within a linear time-invariant framework. We then review the core graph-theoretic concepts of matchings and alternating paths that allow us to formally define the problem of budget-preserving union contraction for driver sets in duplex networks.

\subsection{Foundational Concepts}\label{subsec:foundations}

We first define the basic notation used throughout this paper. A summary is provided in Table \ref{tab:notation}. The terms ``node'' and ``vertex'' and ``edge'' and ``link'' are used interchangeably.

We adopt the multilayer terminology of Kivelä et al.\ and De Domenico et al.: a duplex is a two‑layer multiplex sharing a (possibly aligned) node set; unless stated otherwise, layers differ only in intra‑layer edges while the node set is common \cite{kivela_multilayer_2014, de_domenico_structural_2015}.

\begin{table}[ht]
\caption{Summary of Key Notations and Acronyms.}
\label{tab:notation}
\centering
\resizebox{\linewidth}{!}{%
\begin{tabular}{ll}
\toprule
\textbf{Symbol/Acronym} & \textbf{Description} \\
\midrule
\multicolumn{2}{l}{\textit{\textbf{A. Foundational Concepts}}} \\
$G=(V,E)$ & Directed graph with nodes $V$ and edges $E$. \\
$N$ & Number of nodes, $|V|$. \\
$\mathcal{B}$ & Bipartite representation of $G$. \\
$M, M^*$ & Matching, maximum matching. \\
$D(M)$ & Driver set for a matching $M$. \\
LTI & Linear Time-Invariant. \\
MDS & Minimum Driver Set. \\
\\
\multicolumn{2}{l}{\textit{\textbf{B. Duplex Network Formulation}}} \\
$k_\ell$ & Driver budget for layer $\ell$. \\
$\mathcal{M}_\ell(k_\ell)$ & Feasible matching set for layer $\ell$. \\
$U(M_1, M_2)$ & \UDSfull\ (\UDSabbr), $D_1 \cup D_2$. \\
$U^\star(M_1, M_2)$ & \MinUDSfull\ (\MinUDSabbr), Minimum $D_1 \cup D_2$. \\
CDS & Consistently Driven Set, $D_1 \cap D_2$. \\
CMS & Consistently Matched Set, $(V \setminus D_1) \cap (V \setminus D_2)$. \\
$\text{DD}_\ell$ & Difference-Driver Set for layer $\ell$. \\
$\Delta(M_1, M_2)$ & Difference mass, $|\text{DD}_1| + |\text{DD}_2|$. \\
\\
\multicolumn{2}{l}{\textit{\textbf{C. Algorithms}}} \\
\algoCLAPS\ & Main method: Shortest \CLAPabbr Search. \\
\algoCLAPG\ & Greedy Segment / 1‑\CLAPabbr (fast local search). \\
\algoRSU\,(K) & Random Sample \& Union baseline with K samples per layer. \\
\algoILP\ & Exact ILP solver (gold standard). \\
\\
\multicolumn{2}{l}{\textit{\textbf{D. Experimental Metrics}}} \\
$\Delta N_D$ & Absolute \UDSabbr\ Reduction. \\
$\Delta N_D^{\text{opt}}$ & Optimization Gain over baseline. \\
$R_{opt}$ & Relative Optimization Rate. \\
$\overline{h}$ & Average \CLAPabbr length. \\
ER, BA & Erd\H{o}s-R\'enyi, Barab\'asi-Albert (network models) \cite{erdos_renyi_1959, barabasi_albert_1999}. \\
\bottomrule
\end{tabular}%
}
\end{table}

The dynamics of many networked systems can be modeled using a linear time-invariant (LTI) framework. For a network of $N$ nodes, the state of the system is described by a vector $\mathbf{x}(t) \in \mathbb{R}^{N}$, and its evolution is governed by:
\begin{equation}
\frac{d\mathbf{x}(t)}{dt} = A\mathbf{x}(t) + B\mathbf{u}(t)
\label{eq:lti_system}
\end{equation}
where $A \in \mathbb{R}^{N \times N}$ is the system's interaction matrix and $\mathbf{u}(t) \in \mathbb{R}^{M}$ is the vector of $M$ control inputs applied to a specific set of \emph{driver nodes} via the input matrix $B \in \mathbb{R}^{N \times M}$.

In many complex systems, the exact weights of interactions are unknown. This motivates the use of \emph{structural controllability} \cite{ching-tai_lin_structural_1974}, which depends only on the network's topology. The structural controllability of a network, represented by a directed graph $G=(V, E)$, is determined by the graph's matching properties in its bipartite representation \cite{ching-tai_lin_structural_1974, liu_controllability_2011}.

\begin{definition}[Bipartite Representation]\label{def:bipartite}
For a directed graph $G=(V, E)$, its \textbf{bipartite representation} is a bipartite graph $\mathcal{B}=(V^{+} \cup V^{-}, E_{\mathcal{B}})$. The two disjoint vertex sets are copies of $V$: $V^{+} = \{v^{+} \mid v \in V\}$ and $V^{-} = \{v^{-} \mid v \in V\}$. For every directed edge $(u,v) \in E$, there is a corresponding edge $(u^{+}, v^{-}) \in E_{\mathcal{B}}$.
This construction is the standard reduction used in structural controllability \cite{liu_controllability_2011}.
\end{definition}

A \emph{matching} $M$ in $\mathcal{B}$ is a subset of $E_{\mathcal{B}}$ where no two edges share a common vertex. The set of vertices incident to an edge in $M$ is the saturated vertex set $V_M$.

\begin{definition}[Driver Set]\label{def:driver_set}
For a given matching $M$, the corresponding \textbf{driver set} is the set of nodes in $V$ whose $V^{-}$ counterparts are unmatched in $\mathcal{B}$. Formally,
\begin{equation}\label{eq:driver_def}
D(M) = \{\, v\in V:\ v^{-} \notin V_M \,\}.
\end{equation}
The cardinality of the driver set is determined by the cardinality of the matching, following the identity $|D(M)| = N - |M|$ \cite{liu_controllability_2011, zhang_input_2016}.
\end{definition}

\begin{figure}[!t]
    \centering
    \includegraphics[width=\linewidth]{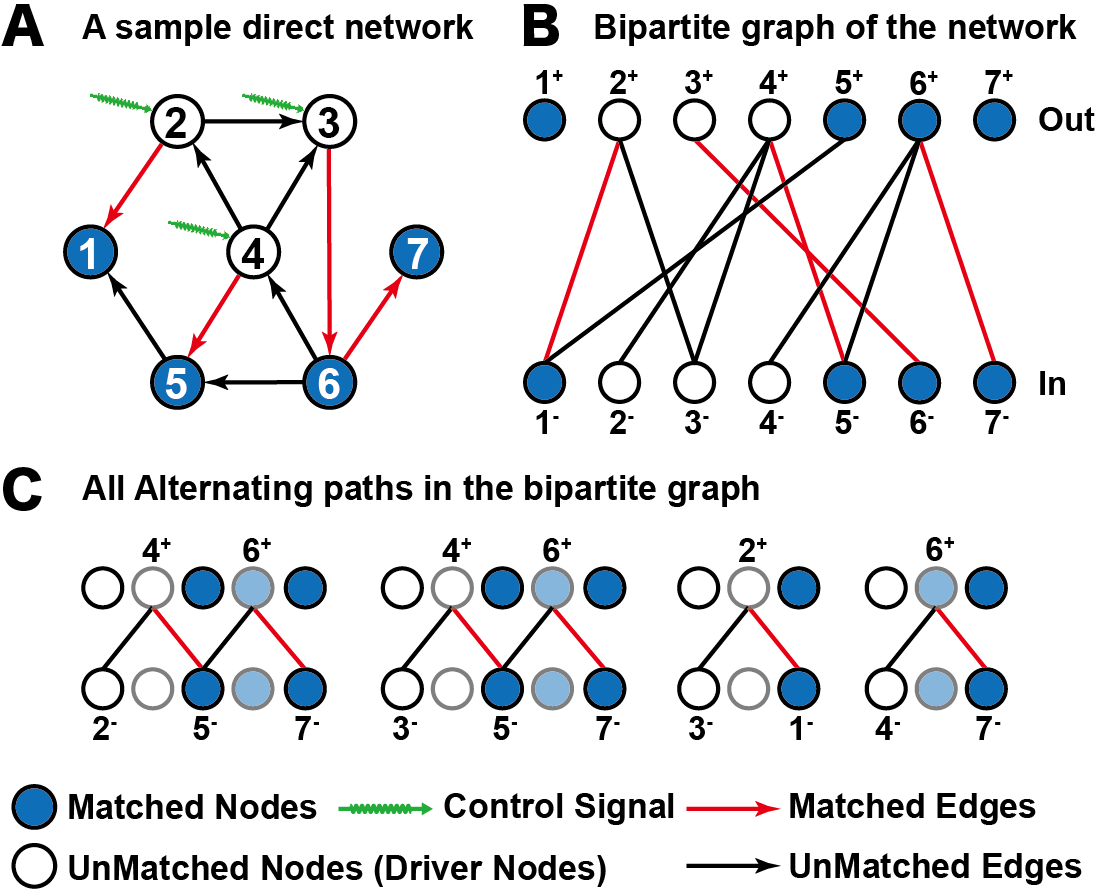}
    \caption{Illustration of foundational concepts for structural controllability. (A) A sample directed network $G=(V,E)$. (B) The bipartite representation $\mathcal{B}$ of the network. A maximum matching $M^*$ is shown with red edges. The vertices in $V^{-}$ that are unmatched by $M^*$ (white circles) correspond to the nodes in the Minimum Driver Set (MDS), which is $D(M^*) = \{2, 3, 4\}$. (C) Examples of $M^*$-alternating paths. Such paths allow for the transformation between different maximum matchings, yielding different MDS configurations of the same minimum size.}
    \label{fig:concepts}
\end{figure}

A Minimum Driver Set (MDS) is a driver set of the smallest possible cardinality, required to ensure structural controllability. The size of an MDS is determined by a \emph{maximum matching} $M^*$, which is a matching of the largest possible size. The Minimum Input Theorem \cite{liu_controllability_2011} states that the size of an MDS is $N_D = N - |M^*|$, and any driver set $D(M^*)$ derived from a maximum matching is an MDS. Fig.\ref{fig:concepts} illustrates these foundational concepts.

\subsection{The Driver Exchange Principle}\label{subsec:driver_exchange}

The configuration of a driver set is not unique, even when its size is fixed. The mechanism for transitioning between these configurations is the $M$-alternating path.

\begin{definition}[Alternating Path]\label{def:alternating_path}
Given a matching $M$, a simple path in $\mathcal{B}$ is \textbf{$M$-alternating} if its edges alternate between belonging to $M$ and not belonging to $M$.
\end{definition}

An $M$-alternating path can be used to modify a matching $M$ into a new matching $M'$ of the same size. This operation, which forms the basis of our approach, allows for the reconfiguration of the driver set while preserving its cardinality. \cite{zhang_structural_2014, zhang_input_2016, zhang_efficient_2017}

\begin{theorem}[Driver Exchange Principle (\emph{cf.} \cite{zhang_structural_2014, zhang_input_2016, zhang_efficient_2017})]\label{thm:exchange}
Let $M$ be a matching. Let $s \in D(M)$ be a driver node and $t \in V \setminus D(M)$ be a non-driver node. There exists an $M$-alternating path $p$ from $s^{-}$ to $t^{-}$ in $\mathcal{B}$ if and only if the matching $M' = M \triangle E(p)$, where $E(p)$ is the set of edges in $p$, satisfies
\begin{enumerate}\itemsep 2pt
    \item[\textnormal{(i)}] $|M'| = |M|$, which implies $|D(M')| = |D(M)|$, and
    \item[\textnormal{(ii)}] $D(M') = (D(M) \setminus \{s\}) \cup \{t\}$.
\end{enumerate}
\end{theorem}

\begin{IEEEproof}
The symmetric difference operation along an $M$-alternating path that connects an unmatched vertex $s^{-}$ to a matched vertex $t^{-}$ preserves the matching property and its cardinality. The operation changes the matching status only at the endpoints, making $s^{-}$ matched and $t^{-}$ unmatched. This corresponds directly to the stated update of the driver set.
\end{IEEEproof}

\subsection{Problem Formulation: Budget-Preserving Union Contraction}\label{subsec:problem_formulation}

We now consider a duplex network, composed of two layers $G_1=(V,E_1)$ and $G_2=(V,E_2)$ on the same set of $N$ nodes. From an engineering perspective, it is often necessary to control each layer with a fixed number of drivers, determined by cost or physical constraints. Let these numbers be the driver budgets $k_1$ and $k_2$. Typically, these budgets correspond to the minimum required for each layer, $k_\ell = N - |M_\ell^*|$, where $M_\ell^*$ is a maximum matching for layer $\ell$. Our problem is not to find these minimum budgets, but to work within them.

\begin{definition}[Fixed-Budget Feasible Set]\label{def:feasible_set}
For each layer $\ell \in \{1,2\}$, given a driver budget $k_\ell$, the \textbf{feasible set of matchings} is
\begin{equation}
\mathcal{M}_\ell(k_\ell) = \{ M_\ell : |M_\ell| = N - k_\ell \}.
\end{equation}
The search space for our problem is the Cartesian product $\mathcal{M}_1(k_1) \times \mathcal{M}_2(k_2)$. A state is a pair $(M_1, M_2)$ from this space.
\end{definition}

For any state $(M_1, M_2)$, the driver sets are $D_1(M_1)$ and $D_2(M_2)$. When a single physical actuator system is used, the total set of required drivers is their union, which we aim to contract.

\begin{definition}[\UDSfull]\label{def:uds}
For a state $(M_1, M_2) \in \mathcal{M}_1(k_1) \times \mathcal{M}_2(k_2)$, the \textbf{\UDSfull\ (\UDSabbr)} is
\begin{equation}
U(M_1, M_2) = D_1(M_1) \cup D_2(M_2).
\end{equation}
\end{definition}

\begin{figure}[!t]
    \centering
    \includegraphics[width=\linewidth]{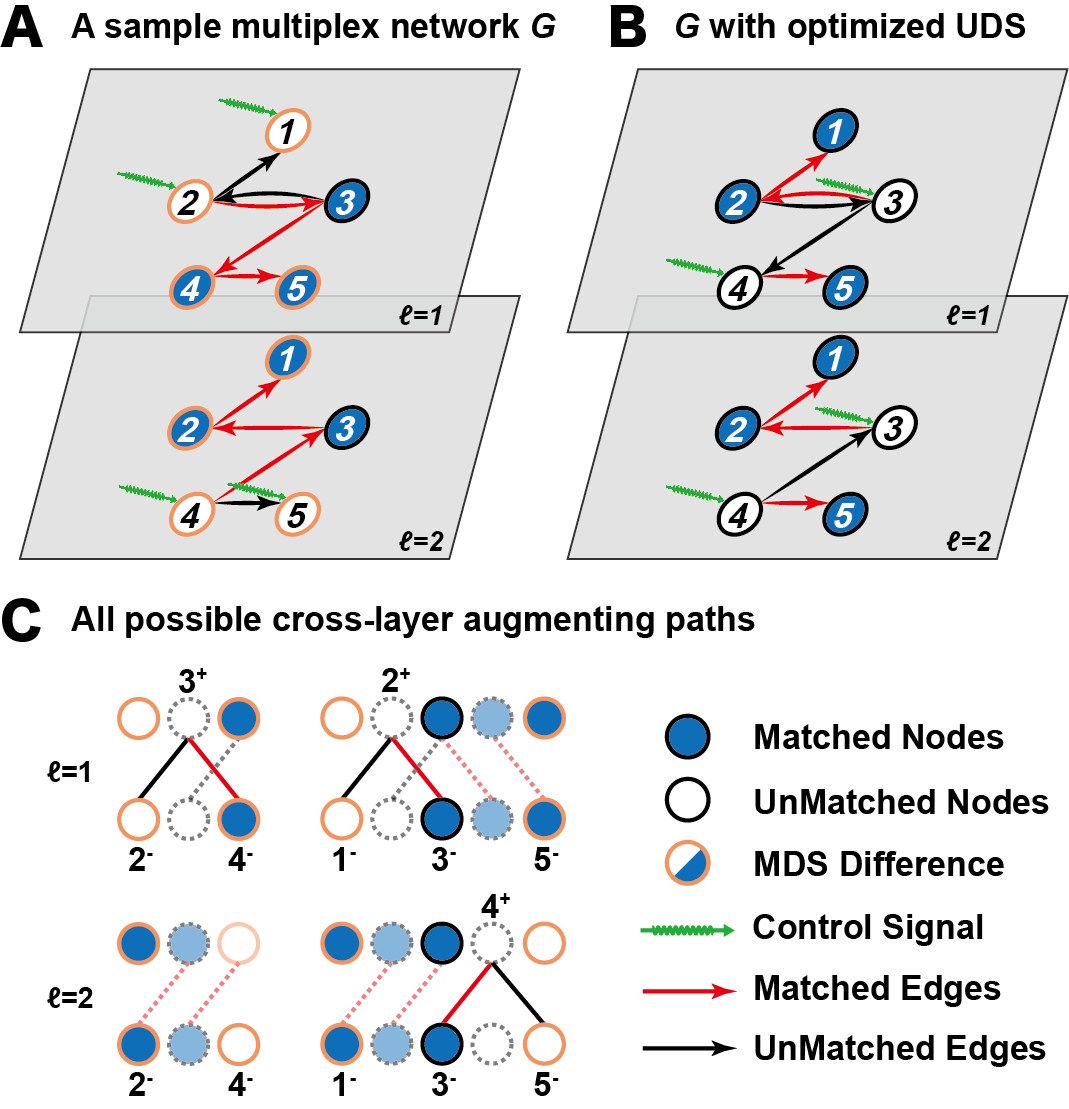}
    \caption{Illustration of the budget-preserving \UDSabbr\ contraction problem. (A) An initial state of a duplex network with driver budgets $k_1=2, k_2=2$. The chosen matchings yield driver sets $D_1=\{1,2\}$ and $D_2=\{4,5\}$. The initial \UDSabbr\ is $\{1,2,4,5\}$, with size 4. (B) A contracted state for the same network. By reconfiguring the matchings within the same budgets, the new driver sets are $D_1'=\{3,4\}$ and $D_2'=\{3,4\}$. The resulting \UDSabbr\ is $\{3,4\}$, with size 2. This represents a more efficient alignment of driver nodes.}
    \label{fig:duplex_opt}
\end{figure}

The choice of matchings within the feasible set critically impacts the size of the \UDSabbr\, as illustrated in Fig.\ref{fig:duplex_opt}. An uncoordinated selection of matchings can lead to a large \UDSabbr\, while a careful alignment can significantly contract it.

We formally define our central problem as follows.

\begin{problem}[Budget-Preserving \UDSabbr\ Contraction]\label{prob:uds_contraction}
Given a duplex network with fixed driver budgets $(k_1, k_2)$ for its layers $(G_1, G_2)$, find a state $(M_1^*, M_2^*)$ in the feasible set $\mathcal{M}_1(k_1) \times \mathcal{M}_2(k_2)$ that solves:
\begin{equation}
(M_1^\star, M_2^\star)
= \underset{\substack{M_1 \in \mathcal{M}_1(k_1)\\ M_2 \in \mathcal{M}_2(k_2)}}{\arg\min}\;
\bigl| D_1(M_1) \cup D_2(M_2) \bigr|.
\label{eq:problem_def}
\end{equation}
\end{problem}

\begin{definition}[\MinUDSfull]\label{def:minuds}
Over the feasible set $\mathcal M_1(k_1)\times\mathcal M_2(k_2)$, the optimal union is
\(
  U^\star \in \arg\min_{(M_1,M_2)} |U(M_1,M_2)|.
\)
We denote its size by $|U^\star|$.
\end{definition}

The primary challenge is that the feasible set $\mathcal{M}_1(k_1) \times \mathcal{M}_2(k_2)$ can be combinatorially large, rendering a brute-force search infeasible. Our work introduces an efficient, path-based algorithm to navigate this space.

\begin{remark}[Protocol for harmonization and budgets.]\label{rem:protocol-harmonization}
We fix a common node set $V$ as the union over layers; nodes isolated in a layer are counted as drivers for that layer. Budgets are set to $k_\ell=N-|M_\ell^*|$ and remain fixed during optimization.
\end{remark}

\subsection{An Equivalent Objective: Minimizing the Difference Mass}\label{subsec:equiv_objective}

To systematically reduce the size of the \UDSabbr\, we partition the node set $V$ based on a given state $(M_1, M_2)$. Let $D_1 = D_1(M_1)$ and $D_2 = D_2(M_2)$.

\begin{definition}[Node Partition]\label{def:partition}
The node set $V$ is partitioned into four disjoint subsets:
\begin{itemize}
    \item The Consistently Driven Set (CDS): $\text{CDS} = D_1 \cap D_2$.
    \item The Consistently Matched Set (CMS): $\text{CMS} = (V \setminus D_1) \cap (V \setminus D_2)$.
    \item The Layer-1 Difference-Driver Set: $\text{DD}_1 = D_1 \setminus D_2$.
    \item The Layer-2 Difference-Driver Set: $\text{DD}_2 = D_2 \setminus D_1$.
\end{itemize}
\end{definition}

Nodes in $\text{DD}_1$ and $\text{DD}_2$ are the source of inefficiency in the \UDSabbr. We quantify this with the following measure.

\begin{definition}[Difference Mass]\label{def:diff_mass}
The \textbf{difference mass} of a state $(M_1, M_2)$ is defined as
\begin{equation}
\Delta(M_1, M_2) = |\text{DD}_1| + |\text{DD}_2|.
\end{equation}
\end{definition}

The size of the \UDSabbr\ is directly related to the difference mass. This relationship transforms the goal of contracting the union into the equivalent goal of reducing the difference mass.

\begin{proposition}[Equivalence of Objectives]\label{prop:equiv_objective}
For any state $(M_1, M_2) \in \mathcal{M}_1(k_1) \times \mathcal{M}_2(k_2)$, the size of the \UDSabbr\ is given by
\begin{equation}\label{eq:union_in_delta}
|U(M_1, M_2)| = \frac{k_1 + k_2 + \Delta(M_1, M_2)}{2}.
\end{equation}
\end{proposition}

\begin{IEEEproof}
By Definition \ref{def:partition}, we have $|D_1| = |\text{CDS}| + |\text{DD}_1| = k_1$ and $|D_2| = |\text{CDS}| + |\text{DD}_2| = k_2$. The size of the \UDSabbr\ is $|U| = |\text{CDS}| + |\text{DD}_1| + |\text{DD}_2|$. Summing the first two identities gives $2|\text{CDS}| + \Delta(M_1, M_2) = k_1 + k_2$. Substituting $|\text{CDS}| = |U| - \Delta(M_1, M_2)$ yields $2(|U| - \Delta(M_1, M_2)) + \Delta(M_1, M_2) = k_1 + k_2$, which simplifies to $2|U| - \Delta(M_1, M_2) = k_1 + k_2$. Rearranging this equation gives the desired result. Since $k_1$ and $k_2$ are fixed budgets, minimizing $|U|$ is equivalent to minimizing $\Delta$.
\end{IEEEproof}

\section{The \CLAPfull\ (\CLAPabbr) Framework}\label{sec:clap_framework}

The Driver Exchange Principle (Theorem \ref{thm:exchange}) provides a mechanism to reconfigure the driver set of a single layer while preserving its budget. To achieve a coordinated contraction of the \UDSabbr\ across two layers, we introduce the \CLAPfull\ (\CLAPabbr), a structured sequence of these single-layer exchanges. A \CLAPabbr\ is designed to convert one node from the Layer-1 Difference-Driver Set ($\text{DD}_1$) and one from the Layer-2 Difference-Driver Set ($\text{DD}_2$) into consistently classified nodes, thereby reducing the difference mass $\Delta$ and contracting the \UDSabbr.

\subsection{Definition of a \CLAPabbr}\label{subsec:clap_definition}

A \CLAPabbr\ is constructed from a sequence of admissible segments, where each segment represents a valid driver exchange in one of the layers.

\begin{definition}[Admissible Segment]\label{def:admissible_segment}
Given a state $(M_1, M_2)$, an ordered pair of nodes $(u,v)$ forms an \textbf{admissible segment in layer $\ell$}, denoted $(u \xrightarrow{\ell} v)$, if there exists an $M_\ell$-alternating path from one endpoint to the other and the nodes satisfy the following polarity conditions:
\begin{itemize}
    \item For layer $\ell=1$: $u \in D_1(M_1)$ and $v \notin D_1(M_1)$. The underlying alternating path enables a driver exchange that removes $u$ from $D_1$ and adds $v$.
    \item For layer $\ell=2$: $u \notin D_2(M_2)$ and $v \in D_2(M_2)$. The underlying alternating path enables a driver exchange that removes $v$ from $D_2$ and adds $u$.
\end{itemize}
The underlying $M_\ell$-alternating path is called a \emph{witness path} for the segment.
\end{definition}

\begin{figure*}[!t]
    \centering
    \includegraphics[width=\linewidth]{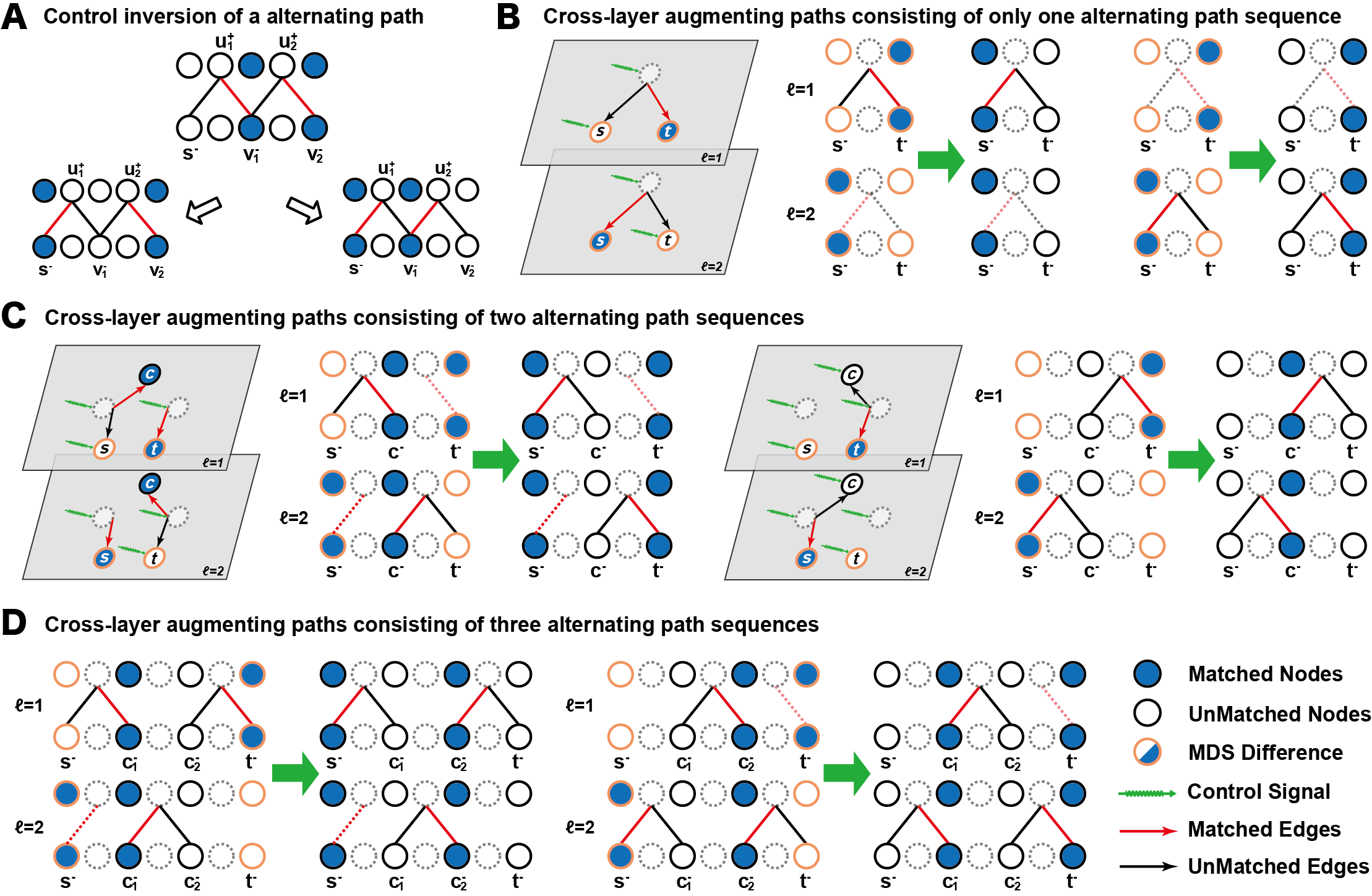}
    \caption{A schematic of a \CLAPfull\ (\CLAPabbr). The path starts at a node $v_0 \in \text{DD}_1$ and ends at $v_k \in \text{DD}_2$. It proceeds via a sequence of admissible segments through relay nodes. In this example of a length-3 \CLAPabbr\, the first segment $(v_0 \xrightarrow{1} v_1)$ operates in layer 1. The relay node $v_1$ must be in CMS to enable the subsequent layer-2 segment $(v_1 \xrightarrow{2} v_2)$. The second relay $v_2$ must be in CDS to enable the final layer-1 segment $(v_2 \xrightarrow{1} v_k)$. Applying this \CLAPabbr\ converts both $v_0$ and $v_k$ into consistently classified nodes, reducing the difference mass $\Delta$ by 2.}
    \label{fig:clap_schematic}
\end{figure*}

By chaining these segments together in an alternating fashion across layers, we define a \CLAPabbr\. A visual representation is provided in Fig.\ref{fig:clap_schematic}.

\begin{definition}[\CLAPfull\ (\CLAPabbr)]\label{def:clap}
A \textbf{\CLAPfull} is a sequence of admissible segments
\[
\mathcal{P}
= \bigl(
  v_0 \xrightarrow{\ell_1} v_1 \xrightarrow{\ell_2} v_2 \xrightarrow{\ell_3} \cdots \xrightarrow{\ell_k} v_k
  \bigr).
\]
that satisfies the following properties:
\begin{enumerate}\itemsep 2pt
    \item[\textnormal{(C1)}] \textbf{Endpoints:} The path starts at a node $v_0 \in \text{DD}_1$ and terminates at a node $v_k \in \text{DD}_2$.
    \item[\textnormal{(C2)}] \textbf{Alternating Layers:} The layers of consecutive segments alternate, i.e., $\ell_{i+1} \neq \ell_i$ for all $i \in \{1, \dots, k-1\}$.
    \item[\textnormal{(C3)}] \textbf{Distinct Nodes:} The nodes in the sequence $(v_0, v_1, \dots, v_k)$ are all distinct.
\end{enumerate}
The intermediate nodes $\{v_1, \dots, v_{k-1}\}$ are called \emph{relay nodes}.
\end{definition}

\begin{remark}[Layer symmetry of \CLAPabbr s]\label{rem:layer-symmetry}
Definition~\ref{def:clap} does not constrain the initial layer: a \CLAPabbr\ may start in layer~1 or in layer~2.
All structural statements used later—such as the Relay Type Constraint (Lemma~\ref{lem:relay_type}),
the \CLAPabbr\ Gain Theorem (Theorem~\ref{thm:clap_gain}), the feasibility of shortest \CLAPabbr s,
and the no-\CLAPabbr\ certificate/minimality result—are invariant under swapping the layer indices
$1\leftrightarrow 2$. When convenient, we may state proofs “assuming the first segment is in layer~1”
purely as a notational convention; no generality is lost. Formally, the index-swap
$(M_1,D_1,\mathrm{DD}_1)\leftrightarrow(M_2,D_2,\mathrm{DD}_2)$ preserves the objective
$|U|=\tfrac{k_1+k_2+\Delta}{2}$ and the decrement claims for \CLAPabbr s, so every assertion has a verbatim
counterpart with layers exchanged.
\end{remark}

The structure of a \CLAPabbr\ imposes strong constraints on the types of nodes that can serve as relays, as established by the following lemma.

\begin{lemma}[Relay Type Constraint]\label{lem:relay_type}
For any relay node $v_i$ ($1 \le i \le k-1$) in a \CLAPabbr\, its classification is determined by the layer of the segment arriving at it:
\begin{itemize}
    \item If the segment $(v_{i-1} \xrightarrow{1} v_i)$ is in layer 1, then the relay node $v_i$ must be in the Consistently Matched Set (CMS).
    \item If the segment $(v_{i-1} \xrightarrow{2} v_i)$ is in layer 2, then the relay node $v_i$ must be in the Consistently Driven Set (CDS).
\end{itemize}
\end{lemma}

\begin{IEEEproof}
Consider a relay $v_i$ that is the endpoint of a layer-1 segment $(v_{i-1} \xrightarrow{1} v_i)$. By Definition \ref{def:admissible_segment}, $v_i \notin D_1$. Since layers must alternate, the next segment is $(v_i \xrightarrow{2} v_{i+1})$, which requires $v_i \notin D_2$. Together, $v_i \notin D_1$ and $v_i \notin D_2$ imply $v_i \in \text{CMS}$. A symmetric argument shows that a relay following a layer-2 segment must be in CDS.
\end{IEEEproof}

\subsection{Monotonic Improvement via \CLAPabbr\ Application}\label{subsec:clap_gain}

Applying a \CLAPabbr\ involves executing the driver exchanges corresponding to each of its segments. This is performed via a batch symmetric difference operation on the matchings of each layer, using the witness paths for all segments. For this process to be well-defined, we require a \emph{feasible} \CLAPabbr, one whose witness paths within the same layer are edge-disjoint. As we will see in Section \ref{sec:algorithm}, the shortest \CLAPabbr s found by our algorithm are guaranteed to be feasible.

The application of a feasible \CLAPabbr\ is a budget-preserving operation that provides a guaranteed, monotonic improvement to our objective function, the difference mass $\Delta$.

\begin{theorem}[\CLAPabbr\ Gain Theorem]\label{thm:clap_gain}
Let $\mathcal{P}$ be a feasible \CLAPabbr\ from $v_0 \in \text{DD}_1$ to $v_k \in \text{DD}_2$ in a state $(M_1, M_2)$. Applying $\mathcal{P}$ yields a new state $(M_1', M_2')$ that satisfies:
\begin{enumerate}\itemsep 2pt
    \item[\textnormal{(i)}] \textbf{Budget Preservation:} The driver budgets are preserved, i.e., $|D_1(M_1')|=k_1$ and $|D_2(M_2')|=k_2$.
    \item[\textnormal{(ii)}] \textbf{Difference Mass Reduction:} The difference mass strictly decreases by two:
    \begin{equation}
    \Delta(M_1', M_2') = \Delta(M_1, M_2) - 2.
    \end{equation}
    \item[\textnormal{(iii)}] \textbf{\UDSabbr\ Contraction:} The size of the Universal Driver Set strictly decreases by one:
    \begin{equation}
    |U(M_1', M_2')| = |U(M_1, M_2)| - 1.
    \end{equation}
\end{enumerate}
\end{theorem}

\begin{IEEEproof}
(i) Each admissible segment within the \CLAPabbr\ corresponds to a driver exchange operation as defined in Theorem \ref{thm:exchange}. This operation is budget-preserving for its layer. Since the application of the \CLAPabbr\ is simply a collection of these operations, the overall budgets $k_1$ and $k_2$ remain invariant.

(ii) We analyze the change in node classifications.
\textbf{Start Node $v_0$:} Initially, $v_0 \in \text{DD}_1$, so $v_0 \in D_1$ and $v_0 \notin D_2$. The first segment of the \CLAPabbr, $(v_0 \xrightarrow{\ell_1} v_1)$, alters the status of $v_0$ in layer $\ell_1$. If $\ell_1=1$, $v_0$ is removed from $D_1$. If $\ell_1=2$, $v_0$ is added to $D_2$. In either case, $v_0$ ceases to be in $\text{DD}_1$ and becomes a consistently classified node (CMS or CDS).
\textbf{End Node $v_k$:} Initially, $v_k \in \text{DD}_2$, so $v_k \notin D_1$ and $v_k \in D_2$. The final segment $(v_{k-1} \xrightarrow{\ell_k} v_k)$ alters the status of $v_k$ in layer $\ell_k$. A symmetric analysis shows that $v_k$ also becomes a consistently classified node.
\textbf{Relay Nodes $v_i$:} By Lemma \ref{lem:relay_type}, each relay node is initially consistently classified (either CMS or CDS). Each relay $v_i$ is an endpoint for two segments, $(v_{i-1} \xrightarrow{\ell_i} v_i)$ and $(v_i \xrightarrow{\ell_{i+1}} v_{i+1})$, where $\ell_{i+1} \neq \ell_i$. The first segment flips its driver status in layer $\ell_i$, and the second flips its status in layer $\ell_{i+1}$. The net effect is that its classification changes from CMS to CDS or vice-versa, but it remains consistently classified.
\textbf{Other Nodes:} Nodes not part of the \CLAPabbr\ are unaffected.
Therefore, the application of the \CLAPabbr\ removes exactly two nodes, $v_0$ and $v_k$, from the difference sets ($\text{DD}_1 \cup \text{DD}_2$) and creates no new ones. This results in a net decrease of 2 in the difference mass $\Delta$.

(iii) This follows directly from Proposition \ref{prop:equiv_objective}. Since $|U| = (k_1 + k_2 + \Delta)/2$ and the budgets $k_1, k_2$ are constant, a decrease of 2 in $\Delta$ leads to a decrease of 1 in $|U|$.
\end{IEEEproof}

The \CLAPabbr\ Gain Theorem is central to our method. It guarantees that every successful \CLAPabbr\ operation makes provable, monotonic progress toward contracting the \UDSabbr\, forming the basis of our iterative algorithm.

\subsection{Shortest \CLAPabbr s are Feasible}\label{subsec:shortest-feasible}
We call a \CLAPabbr\ \emph{feasible} if, for each layer, the witness $M_\ell$-alternating paths of all its layers-$\ell$ segments are edge-disjoint, so that the batched symmetric difference is well-defined.

\begin{lemma}[Same-layer overlap implies compressibility]\label{lem:overlap-compress}
Fix $\ell\in\{1,2\}$ and let
$\mathcal{P}=((v_{0}\xrightarrow{\ell_{1}}v_{1}),\dots,(v_{k-1}\xrightarrow{\ell_{k}}v_{k}))$
be a \CLAPabbr\ of minimum length among all \CLAPabbr s with endpoints $(v_0,v_k)$.
If there exist $i<j$ with $\ell_{i}=\ell_{j}=\ell$ such that the two layer-$\ell$ witness alternating paths overlap on at least one edge, then there exists another \CLAPabbr\ between $(v_0,v_k)$ with strictly fewer segments.
\end{lemma}

\begin{corollary}[Shortest \CLAPabbr s are feasible]\label{cor:shortest-feasible}
Every shortest \CLAPabbr\ admits a choice of witness alternating paths that are edge-disjoint within each layer; consequently, the batched symmetric-difference operation is well-defined.
\end{corollary}

\noindent Proofs are deferred to the Supplementary Materials.

\subsection{Existence via a Layer-Labeled Meta-Graph}\label{subsec:meta_graph_main}
For each layer $\ell\in\{1,2\}$, let $H_\ell$ be the symmetric-difference graph $(V^+\!\cup V^-,\,M_\ell\triangle\widehat M_\ell)$ between the current matching $M_\ell$ and any comparator matching $\widehat M_\ell$ of the same size. Each connected component of $H_\ell$ is an even cycle or an alternating path with $V^-$-endpoints.

\begin{definition}[Layer-labeled meta-graph]\label{def:layer-meta}
Define an undirected multigraph $\mathcal{K}=(V,\mathcal{E})$ by adding, for every path component of $H_\ell$ with $V^-$ endpoints $\{x^-,y^-\}$, an undirected edge $\{x,y\}$ labeled by $\ell$. (Even cycles add no edge.)
\end{definition}

\begin{lemma}[Degrees and alternating labels]\label{lem:deg-alt}
Every connected component of $\mathcal K$ is a simple path or a cycle. Moreover, along any path component, the edge labels must alternate $1,2,1,2,\dots$.
\end{lemma}

We orient $\mathcal K$ by declaring exactly one admissible direction on each labeled edge:

\begin{definition}[Admissible orientation]\label{def:admissible-orient-layer}
For $e=\{u,v\}\in\mathcal E$ with label $\lambda(e)=\ell$, orient $e$ as $(u\xrightarrow{\ell} v)$ iff $u\in D_\ell$, $v\notin D_\ell$, and $v$ is $M_\ell$-alternating reachable from $u$; otherwise, orient it as $(v\xrightarrow{\ell} u)$. Denote the result by $\overrightarrow{\mathcal K}$.
\end{definition}

\begin{lemma}[Well-defined admissible orientation]\label{lem:well-defined-orient}
For each labeled edge $e=\{u,v\}$, exactly one of $(u\xrightarrow{\ell} v)$ or $(v\xrightarrow{\ell} u)$ is admissible in the current state.
\end{lemma}

\begin{lemma}[Meta-to-\CLAPabbr\ translation]\label{lem:meta-to-clap}
Any directed label-alternating path in $\overrightarrow{\mathcal K}$ from a node in $\mathrm{DD}_1$ to a node in $\mathrm{DD}_2$ induces a \CLAPabbr\ whose segments follow the labels and whose witnesses are alternating paths between the corresponding $V^-$-endpoints in $H_\ell$.
\end{lemma}

To quantify improvement, we classify endpoints and sum their contributions component-wise:

\begin{proposition}[Component-wise contribution to $\Delta$]\label{prop:component-delta}
For any path component of $\mathcal K$, only those with endpoints of types $(\mathsf L,\mathsf R)$ (losing a layer-1 driver and losing a layer-2 driver, respectively, with respect to the comparator state) contribute a strictly positive decrease $+2$ to the difference mass $\Delta$; components with an $\mathsf N_1$ or $\mathsf N_2$ endpoint are neutral, and all other components contribute $\le 0$.
\end{proposition}

\begin{theorem}[Existence of an improving \CLAPabbr]\label{thm:existence-clap}
If there exists a feasible state $(\widehat M_1,\widehat M_2)$ with $\Delta(\widehat M_1,\widehat M_2) < \Delta(M_1,M_2)$, then $\overrightarrow{\mathcal K}$ contains a directed path from $\mathrm{DD}_1$ to $\mathrm{DD}_2$. Consequently, a \CLAPabbr\ exists in the current state.
\end{theorem}

\noindent Proofs are deferred to the Supplementary Materials.

\section{The \algoCLAPS\ Algorithm}\label{sec:algorithm}

We refer to our shortest \CLAPabbr\ guided search as \algoCLAPS. At each iteration, it finds a shortest \CLAPabbr\ and applies its layer-wise alternating-path symmetric differences in batch, which strictly reduces the union size by one (Theorem~\ref{thm:clap_gain}).

\subsection{Algorithm Description}\label{subsec:algorithm_description}

The \algoCLAPS\ algorithm begins with an initial pair of matchings $(M_1, M_2)$ that satisfy the given driver budgets $(k_1, k_2)$. In a typical scenario where the goal is to align MDS, this initialization is performed by computing maximum matchings for each layer. The algorithm then enters an iterative loop. In each iteration, it searches for a shortest \CLAPabbr\ connecting the difference-driver sets $\text{DD}_1$ and $\text{DD}_2$. If such a path is found, it is applied to update the matchings and driver sets, which, by Theorem \CLAPabbr\ Gain Theorem (in the main text), guarantees a reduction in the \UDSabbr\ size. If no \CLAPabbr\ exists, the algorithm terminates, as no further budget-preserving improvement of this type is possible. The high-level procedure is outlined in Algorithm \ref{alg:CLAPS}.\footnote{Reference implementation: \url{https://github.com/njnklab/CLAP-S_Algorithm}.}

\label{alg:CLAPS}
\begin{algorithmic}[1]
\Require Duplex network layers $G_1=(V,E_1)$, $G_2=(V,E_2)$, and driver budgets $k_1, k_2$.
\State Initialize $M_1 \in \mathcal{M}_1(k_1)$ and $M_2 \in \mathcal{M}_2(k_2)$.
\State Compute initial driver sets $D_1 \gets D(M_1)$ and $D_2 \gets D(M_2)$.
\While{\textbf{true}}
  \State $\mathcal{P} \gets \textsc{FindShortestCLAP}(M_1, M_2, D_1, D_2)$ \Comment{Search for a shortest \CLAPabbr.}
  \If{$\mathcal{P}$ is null}
    \State \textbf{break} \Comment{No \CLAPabbr\ found; state is stable.}
  \EndIf
  \State $(M_1, M_2, D_1, D_2) \gets \textsc{ApplyCLAP}(\mathcal{P}, M_1, M_2, D_1, D_2)$ \Comment{Apply the \CLAPabbr.}
\EndWhile
\State \Return $(M_1, M_2, D_1, D_2)$ \Comment{Return the contracted state.}
\end{algorithmic}

The core of the algorithm lies in the \textsc{FindShortestCLAP} subroutine. This is implemented using a layer-alternating Breadth-First Search (BFS). The search begins simultaneously from all nodes in the source set $\text{DD}_1$. The BFS expands layer by layer, exploring admissible segments to valid relay nodes (CMS or CDS, as per Lemma \ref{lem:relay_type}) while alternating between layers. The search terminates as soon as a node in the target set $\text{DD}_2$ is reached. This BFS structure naturally discovers a \CLAPabbr\ with the minimum number of segments. Finding a shortest \CLAPabbr\ is crucial, as it can be proven that any such path is guaranteed to be feasible (i.e., its witness paths in any given layer are edge-disjoint), ensuring that the \textsc{ApplyCLAP} operation is well-defined.

\label{alg:FindCLAP}
\begin{algorithmic}[1]
\Require $M_1,M_2$ (maximum matchings), $D_1,D_2$ (driver sets)
\State $Q \gets \text{empty queue}$
\State $\mathrm{pred},\mathrm{predLayer} \gets \text{empty maps}$ \Comment{for backtracking}
\State $\mathrm{visitedState} \gets \varnothing$ \Comment{states of the form $(\text{node},\text{nextLayer})$}
\ForAll{$s \in \mathrm{DD}_1$}
  \For{$\ell\in\{1,2\}$}
    \State enqueue $(s,\ell)$ into $Q$
    \State $\mathrm{visitedState}\gets \mathrm{visitedState}\cup\{(s,\ell)\}$
    \State $\mathrm{pred}(s,\ell)\gets \text{nil}$;\quad $\mathrm{predLayer}(s,\ell)\gets \text{nil}$
  \EndFor
\EndFor
\While{$Q$ not empty}
  \State $(u,\ell)\gets Q.\text{pop}()$ \Comment{$\ell$ is the layer to expand from $u$}
  \State $R \gets \textsc{AltReach}(u \mid M_\ell)$ \Comment{\emph{destinations only}; excludes $u$}
  \State $T \gets R \cap \mathrm{DD}_2$
  \If{$T \neq \varnothing$}
     \State \Return $\textsc{UnrollSegments}((u,\ell), \text{pick any } v\in T, \mathrm{pred},\mathrm{predLayer})$
  \EndIf
  \State $R_{\text{relay}} \gets (R \cap \mathrm{CMS})$ if $\ell{=}1$ else $(R \cap \mathrm{CDS})$
  \ForAll{$x \in R_{\text{relay}}$}
     \If{$(x,3-\ell)\notin \mathrm{visitedState}$}
        \State $\mathrm{visitedState}\gets\mathrm{visitedState}\cup\{(x,3-\ell)\}$
        \State $\mathrm{pred}(x,3-\ell)\gets (u,\ell)$;\; $\mathrm{predLayer}(x,3-\ell)\gets \ell$
        \State enqueue $(x,3-\ell)$ into $Q$
     \EndIf
  \EndFor
\EndWhile
\State \Return $\varnothing$
\end{algorithmic}

\begin{remark}
    (i) \textsc{MaxMatching} uses Hopcroft--Karp \cite{hopcroft_karp_1973} on the bipartite representation.  
    (ii) \textsc{Drivers} collects unmatched $V^-$-side vertices under $M_\ell$.  
    (iii) \textsc{FindShortestCLAP} is a layer-alternating BFS starting from all $s\in \mathrm{DD}_1$.
    (iv) \textsc{ApplyCLAP} reconstructs layer-internal alternating paths lazily and applies symmetric differences; it updates both $M_\ell$ and $D_\ell$.
    Other auxiliary functions are provided in the supplementary information.
\end{remark}

\subsection{Convergence and Stability Certificate}\label{subsec:convergence_stability}

The design of the \algoCLAPS\ algorithm ensures both convergence and the existence of a clear termination condition that serves as a certificate of irreducibility.

\subsubsection{Convergence.} The algorithm is guaranteed to terminate. By the \CLAPabbr\ Gain Theorem \ref{thm:clap_gain}, each successful application of a \CLAPabbr\ reduces the difference mass $\Delta$, which is a non-negative integer, by exactly 2. Since the initial difference mass $\Delta_0$ is finite (at most $k_1 + k_2$), the loop can run for at most $\Delta_0/2$ iterations before no more \CLAPabbr s can be found.

\subsubsection{\CLAPabbr-Stability.} The algorithm's termination condition provides a powerful certificate. We define a state as "\CLAPabbr-stable" if it admits no \CLAPabbr s.

\begin{definition}[\CLAPabbr-Stable State]\label{def:clap_stable}
A state $(M_1, M_2) \in \mathcal{M}_1(k_1) \times \mathcal{M}_2(k_2)$ is \textbf{\CLAPabbr-stable} if there exists no \CLAPabbr\ from $\text{DD}_1$ to $\text{DD}_2$.
\end{definition}

The main theoretical result of this paper is that such a state is not only a fixed point of our algorithm but also corresponds to a solution to Problem \ref{prob:uds_contraction}.

\begin{theorem}[\CLAPabbr-or-Optimal]\label{thm:no_clap_certificate}
For the current state $(M_1,M_2)\in\mathcal M_1\times\mathcal M_2$, the following are equivalent:
\begin{enumerate}\itemsep 1pt
\item There is no \CLAPabbr\ from $\mathrm{DD}_1$ to $\mathrm{DD}_2$.
\item $\Delta(M_1,M_2)$ is minimal over $\mathcal M_1\times\mathcal M_2$.
\item $|U(M_1,M_2)|$ is minimal over $\mathcal M_1\times\mathcal M_2$.
\end{enumerate}
This optimality certificate is analogous in spirit to Berge's augmenting-path characterization of maximum matchings~\cite{berge_two_1957}.
\end{theorem}

\noindent Proof is deferred to the Supplementary Materials.

\subsubsection{Time Complexity.}
Initialization with Hopcroft--Karp on the two bipartite layers costs
$O(|E_1|\sqrt{|V|} + |E_2|\sqrt{|V|})$ \cite{hopcroft_karp_1973}.
One invocation of \textsc{FindShortestCLAP} consists of $h$ \emph{layer-floods}
(where $h$ is the number of segments of the discovered shortest \CLAPabbr), each being a
multi-source alternating BFS: $O(|E_1|)$ for a flood in layer 1 and $O(|E_2|)$ in layer 2.
Hence one search is $O\!\big(h\cdot (|E_1|+|E_2|)\big)$.

Each successful \CLAPabbr\ reduces $\Delta$ by $2$, so the number of iterations is at most
$\Delta_0/2 \le |V|/2$. Aggregating and adding initialization, we obtain the \emph{worst-case} bound
\begin{align*}
& O\!\Big(|E_1|\sqrt{|V|} + |E_2|\sqrt{|V|} + \Big(\sum_{r=1}^{R} h_r\Big)\,(|E_1|+|E_2|)\Big)\\
&\qquad\le\; O\!\big(|V|^2\, (|E_1|+|E_2|)\big).
\end{align*}
where $R\le |V|/2$ and $h_r$ are the $r$-th \CLAPabbr\ length.
Empirically, on both synthetic and real networks we observe \emph{very short} \CLAPabbr s
($\bar h \approx 1$), for which each search is $O(|E_1|+|E_2|)$ and the total running time is
\begin{align*}
& O\!\big((|E_1|+|E_2|)\sqrt{|V|}\big)
\;+\;
O\!\big(|V|\, (|E_1|+|E_2|)\big).
\end{align*}
which matches our measured runtimes.

\section{Experimental Results and Analysis}
\label{sec:experiments}

To comprehensively evaluate the performance of our proposed \algoCLAPS\ algorithm, we conducted extensive experiments on both synthetically generated networks and a diverse collection of real-world duplex network systems. This section details the experimental setup, including the datasets, the baseline algorithm used for comparison, and the performance metrics.

\subsection{Description of the Network Dataset}
\label{subsec:dataset}

\subsubsection{Synthetic Networks}

To systematically investigate the algorithm's performance under controlled conditions, we generated duplex networks using two canonical models: Erd\H{o}s-R\'enyi (ER) random graphs \cite{erdos_renyi_1959} and Barab\'asi-Albert (BA) scale-free networks \cite{barabasi_albert_1999}. These models allow us to test \algoCLAPS\ across networks with homogeneous and heterogeneous degree distributions \cite{erdos_renyi_1959, barabasi_albert_1999}.

For all synthetic networks, the number of nodes $N$ was fixed at $1000$. We systematically varied the average degree $\langle k \rangle$ from 2.0 to 10.0, enabling the study of networks ranging from sparse to relatively dense topologies. The construction of a duplex network involved first generating a base layer (Layer 1) and then generating a second layer (Layer 2) with a specified Jaccard similarity of their edge sets, allowing us to control the inter-layer structural overlap from low (0.1) to high (0.9). This process yielded duplex ER networks (ER-ER), duplex BA networks (BA-BA), and hybrid duplex networks (ER-BA), providing a robust dataset for evaluating the \algoCLAPS\ algorithm.

\subsubsection{Real-World Networks}

To assess the \algoCLAPS\ algorithm's applicability in practical scenarios, we curated a dataset of real-world systems from a multilayer network collection \cite{de_domenico_structural_2015}. These systems often comprise more than two layers and may have different node sets in each layer. For our experiments, we constructed duplex networks by selecting two functionally significant layers from each system and considering all the nodes present in any layer of the two layers. This ensures a consistent node set $V$ across the duplex structure. The selected networks span various domains. For instance, genetic networks map biological interactions, where we constructed duplex layers representing distinct mechanisms like direct versus physical association interactions from the BioGRID database \cite{stark_biogrid_2006}, or positive versus negative genetic interactions from studies like the Yeast Landscape \cite{costanzo_genetic_2010}. Similarly, neuronal networks include connectome data, such as the \textit{C. elegans} connectome \cite{chen_wiring_2006}, where layers distinguish between electrical junctions and chemical synapses. In the realm of social networks, we analyzed data capturing online interactions, primarily from Twitter \cite{omodei_characterizing_2015, de_domenico_unraveling_2020}, where layers differentiate between user actions such as retweets and mentions. Finally, human relationship networks map interpersonal ties within closed communities \cite{krackhardt_cognitive_1987, raub_emmanuel_2005, coleman_diffusion_1957}, with layers representing different social connections like friendship, advice, or co-working relationships. The properties of these curated real-world duplex networks are summarized in Table~\ref{tab:real_world_networks}.

\begin{table*}[!htbp]
\centering
\caption{Properties of the real-world duplex networks used in experiments. For each system, two representative layers were selected to form a duplex network on their common set of nodes. The node count $N$ represents the size of this common set. The average degree $\langle k \rangle$ is calculated with respect to the nodes present in the given layer.}
\label{tab:real_world_networks}
\resizebox{\textwidth}{!}{%
\begin{tabular}{llcccccc}
\toprule
\textbf{Network Type} & \textbf{Network Name} & \textbf{Nodes ($N$)} & \textbf{Nodes in Layer 1} & \textbf{Nodes in Layer 2} & \textbf{Avg. Degree ($\langle k \rangle$)} & \textbf{$\langle k_1 \rangle$} & \textbf{$\langle k_2 \rangle$} \\
\midrule
\multirow{8}{*}{Genetic \& Neuronal} 
  & Arabidopsis             & 6903   & 5493  & 2859  & 5.29  & 5.05  & 3.09  \\
  & Celegans                & 3191   & 3126  & 239   & 3.68  & 3.56  & 2.62  \\
  & Drosophila              & 8060   & 7356  & 2851  & 9.20  & 6.55  & 9.11  \\
  & HumanHIV1               & 994    & 758   & 380   & 2.62  & 2.29  & 2.28  \\
  & SacchPomb               & 2622   & 971   & 2402  & 7.01  & 3.47  & 6.25  \\
  & Rattus                  & 2593   & 2035  & 1017  & 3.17  & 2.96  & 2.15  \\
  & YeastLandscape          & 4455   & 4422  & 4432  & 85.89 & 30.41 & 55.99 \\
  & CelegansConnectome      & 275    & 253   & 260   & 19.42 & 8.15  & 12.61 \\
\midrule
\multirow{8}{*}{Social} 
  & Cannes                  & 438513 & 340349& 233735& 4.14  & 2.92  & 3.52  \\
  & MLKing                  & 327660 & 288738& 79070 & 2.31  & 2.02  & 2.20  \\
  & MoscowAthletics         & 88777  & 74688 & 46821 & 4.45  & 2.81  & 3.95  \\
  & NYClimate               & 102416 & 94574 & 50054 & 6.75  & 4.52  & 5.26  \\
  & NBAFinals               & 747690 & 690288& 321515& 4.39  & 3.06  & 3.64  \\
  & Sanremo                 & 55871  & 49904 & 34564 & 10.95 & 8.50  & 5.43  \\
  & UCLFinal                & 676654 & 590337& 293682& 3.91  & 3.00  & 2.98  \\
  & GravitationalWaves      & 361846 & 321307& 118282& 3.40  & 2.76  & 2.89  \\
\midrule
\multirow{6}{*}{Human Relationship} 
  & KrackhardtHighTech-f\&a & 21     & 21    & 21    & 27.81 & 18.10 & 9.71  \\
  & KrackhardtHighTech-f\&r & 21     & 21    & 21    & 11.62 & 9.71  & 1.90  \\
  & LazegaLawFirm-f\&a      & 71     & 71    & 69    & 41.32 & 25.13 & 16.67 \\
  & LazegaLawFirm-f\&c      & 71     & 69    & 71    & 47.30 & 16.67 & 31.10 \\
  & PhysiciansInnovation-f\&a & 238   & 215   & 228   & 8.29  & 4.47  & 4.44  \\
  & PhysiciansInnovation-f\&d & 241   & 231   & 228   & 8.89  & 4.89  & 4.44  \\
\bottomrule
\end{tabular}%
}
\end{table*}

\subsection{Baseline Algorithms and Performance Metrics}

To rigorously evaluate the \algoCLAPS\ algorithm, we define key performance metrics focusing on the extent of \UDSabbr\ optimization and computational efficiency.

The primary measure of optimization achieved by an algorithm ($\mathcal{A}$) is the \textbf{Absolute \UDSabbr\ Reduction} compared to an initial, unoptimized state ($\mathcal{I}$). This metric, denoted as $\Delta N_D(\mathcal{A} \mid \mathcal{I})$, quantifies the number of driver nodes saved by algorithm $\mathcal{A}$ relative to the simple union of arbitrarily chosen initial MDSs for each layer ($\text{\UDSabbr}_0$). It is calculated as:
\begin{equation}
\label{eq:deltaND}
\Delta N_D(\mathcal{A} \mid \mathcal{I})
= \bigl|\text{\UDSabbr}\bigr|_0 - \bigl|\text{\UDSabbr}\bigr|_{\mathcal{A}}.
\end{equation}
where $|\text{\UDSabbr}|_0 = |\text{MDS}_1^{\text{initial}} \cup \text{MDS}_2^{\text{initial}}|$ and $|\text{\UDSabbr}|_{\mathcal{A}}$ are the final \UDSabbr\ sizes after applying algorithm $\mathcal{A}$. A larger value of $\Delta N_D$ indicates a greater optimization effect.

Computational efficiency is assessed by comparing the execution times of \algoCLAPS\ and the baseline algorithm.

For a comparative benchmark, we implemented several baselines as follows. 
\subsubsection{\algoRSU\ (Random Sample \& Union).}
The \algoRSU\ algorithm leverages the non-uniqueness of MDSs through random sampling:
\begin{enumerate}
    \item \textbf{Generate MDS Candidates:} For each layer $G_{\alpha}$, a randomized maximum matching algorithm is executed $K$ times, yielding a set of $K$ potentially distinct MDSs, $\mathcal{S}_{\alpha} = \{\text{MDS}_{\alpha,1}, \dots, \text{MDS}_{\alpha,K}\}$.
    \item \textbf{Find Minimum Union:} All $K \times K$ pairwise unions $(\text{MDS}_{1,i} \cup \text{MDS}_{2,j})$ are formed.
    \item \textbf{Return Best Result:} The union yielding the minimum cardinality, $|\text{\UDSabbr}|_{\algoRSU} = \min_{i,j} |\text{MDS}_{1,i} \cup \text{MDS}_{2,j}|$, is selected as the result.
\end{enumerate}
In our experiments, we set $K=20$. While \algoRSU\ explores the solution space, its search is undirected. \algoCLAPS, in contrast, performs a guided search for the optimal solution starting from a single initial matching pair.

\subsubsection{\algoCLAPG\ (Greedy Segment / 1‑\CLAPabbr).}
Starting from $(M_1,M_2)$, repeatedly search for any single admissible segment in either layer that reduces $|U|$ (not necessarily part of a shortest \CLAPabbr) and apply it immediately; stop when no such segment exists.
\algoCLAPG\ applies any single admissible segment (a length‑1 \CLAPabbr\ fragment) that
strictly reduces $|U|$, and repeats until none exists. Although a single segment is not a
full cross‑layer path, we keep the \CLAPabbr\ naming for family consistency. \algoCLAPG\ probes a stronger local search neighborhood than \algoRSU, but lacks \algoCLAPS's global guidance (shortest alternating layer sequence and batched feasibility).

\subsubsection{\algoILP\ (gold standard).}
We formulate a budget-fixed integer program over the bipartite variables of the two layers to minimize $|U|$ subject to $|M_\ell|=N-k_\ell$ and matching constraints, and solve it with a modern CP-SAT/ILP solver \cite{perron_or_tools_2023}.
This baseline serves as ground truth to assess solution quality on small graphs; in all such cases, \algoCLAPS's final $|U|$ matched the ILP optimum in our tests.

\subsubsection{Performance Metrics}
To directly quantify the additional optimization achieved by 's guided search over \algoRSU's random sampling, we define the \textbf{Optimization Gain}, denoted as $\Delta N_{D}^{opt}$. This metric represents the absolute difference in the final \UDSabbr\ sizes achieved by the two algorithms:
\begin{equation}
\Delta N_{D}^{opt} = |\text{\UDSabbr}|_{\algoRSU} - |\text{\UDSabbr}|_{\algoCLAPS}
\label{eq:net_gain}
\end{equation}
A positive value for $\Delta N_{D}^{opt}$ indicates the number of additional driver nodes saved by using \algoCLAPS\ instead of the \algoRSU\ baseline. This is equivalent to the difference in their respective absolute reductions, i.e., $\Delta N_{D}^{opt} = \Delta N_D(\algoCLAPS\ \mid \mathcal{I}) - \Delta N_D(\algoRSU\ \mid \mathcal{I})$.

Finally, to measure the relative improvement, we use the \textbf{Relative Optimization Rate}, denoted as $R_{opt}$, calculated as:
\begin{equation}
\label{eq:Ropt}
R_{\mathrm{opt}}
= \frac{\bigl|\text{\UDSabbr}\bigr|_{\algoRSU} - \bigl|\text{\UDSabbr}\bigr|_{\algoCLAPS}}
       {\bigl|\text{\UDSabbr}\bigr|_{\algoRSU}}
\times 100\%.
\end{equation}
A higher $R_{opt}$ signifies a more substantial optimization achieved by \algoCLAPS. This metric is undefined if $|\text{\UDSabbr}|_{\algoRSU}$ is zero.

\subsubsection{Artifact and Dependencies}
We release a self-contained artifact with our \CLAPabbr-S implementation and all baselines (\algoRSU, \algoCLAPG, \algoILP), including logging and evaluation scripts: \url{https://github.com/njnklab/CLAP-S_Algorithm}. The code is written in Python and relies on \texttt{networkx} for graph operations and, optionally, OR-Tools or PuLP for the ILP baseline \cite{hagberg_networkx_2008, perron_or_tools_2023, mitchell_pulp_2011}. We provide instructions to reproduce every figure and table from a fresh clone (single command per experiment) and list exact package versions in the repository’s manifest. For transparency, we will pin the commit hash in the camera-ready.

All experiments were conducted on a workstation equipped with an Intel\textsuperscript{\textregistered} Core\textsuperscript{TM} i9-14900K processor, featuring 24 cores and 32 threads, with a base frequency of 3.2\,GHz. The system includes 125\,GiB of DDR5 RAM. The operating system was Ubuntu 24.04.2 LTS (codename \textit{noble}) with a 64-bit x86\_64 architecture.

\subsection{Results on Synthetic Networks}
\label{subsec:results_synthetic}

We first evaluate the performance of \algoCLAPS\ on synthetic duplex networks with $N=1000$ nodes, comparing it against the \algoRSU\ baseline and the initial unoptimized state. All results are averaged over 10 independent network instantiations for each parameter configuration to ensure statistical robustness.

\subsubsection{\UDSabbr\ Optimization Count across Network Types and Densities}

\begin{figure*}[!t]
    \centering
    \includegraphics[width=\textwidth]{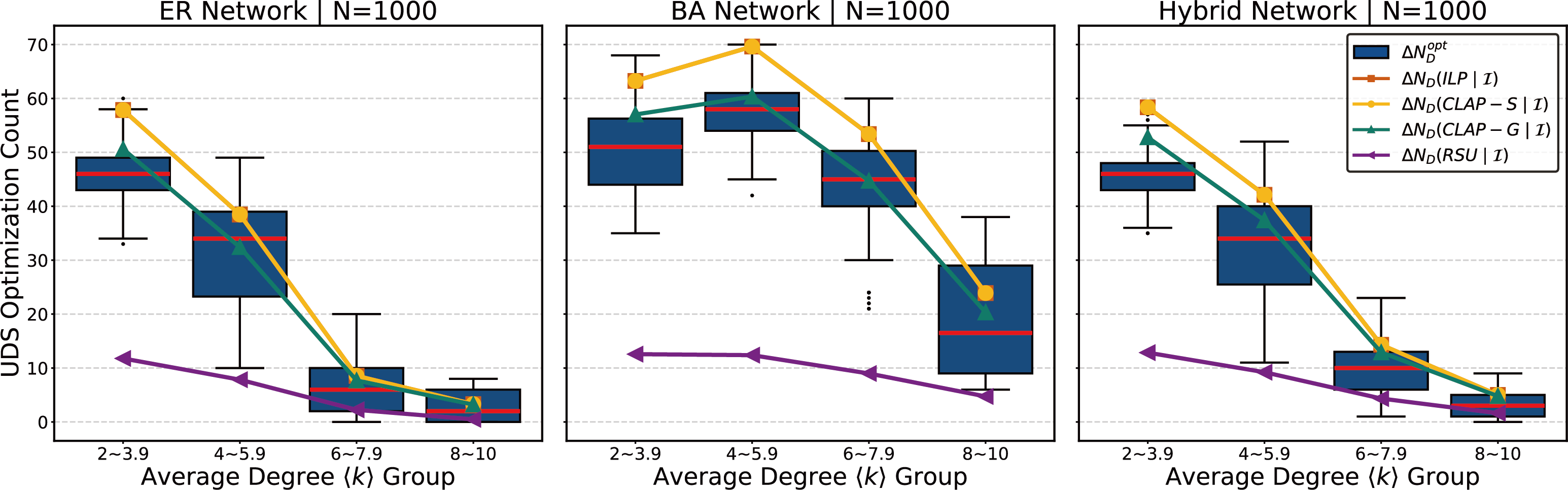}
    \caption{\UDSabbr\ Optimization Count for \algoCLAPS\ and \algoRSU\ on synthetic networks ($N=1000$). The y-axis shows the number of driver nodes saved. Box plots ($\Delta N_{D}^{\text{opt}}$) represent the distribution of the additional nodes saved by \algoCLAPS\ compared to \algoRSU. The orange line with triangles ($\Delta N_D(\algoCLAPS\ \mid \mathcal{I})$) shows the mean optimization by \algoCLAPS\ relative to the initial state. The yellow line with squares ($\Delta N_D(\algoRSU\ \mid \mathcal{I})$) shows the same for \algoRSU. Subplots correspond to ER-ER, BA-BA, and ER-BA (Hybrid) duplex networks, grouped by average degree $\langle k \rangle$.}
    \label{fig:opt_vs_baseline_main}
\end{figure*}

Fig.~\ref{fig:opt_vs_baseline_main} illustrates the optimization performance of \algoCLAPS\ and \algoRSU\ across different network topologies and densities. The results reveal \algoCLAPS's consistent and substantial superiority. The mean number of driver nodes saved by \algoCLAPS\ relative to the initial state, $\Delta N_D(\algoCLAPS\ \mid \mathcal{I})$, is significantly greater than that achieved by \algoRSU, $\Delta N_D(\algoRSU\ \mid \mathcal{I})$, across all tested conditions. Furthermore, the graph optimization gain, $\Delta N_{D}^{\text{opt}}$, which directly measures the additional nodes saved by \algoCLAPS\ compared to \algoRSU, is consistently positive, as shown by the box plots. This confirms that \algoCLAPS's guided search for \CLAPabbr s is fundamentally more effective than \algoRSU's undirected random sampling.

The network density, represented by the average degree $\langle k \rangle$, strongly influences the potential for optimization. For all network types, the absolute number of nodes that can be optimized is highest in sparse regimes and decreases as the networks become denser. This trend is expected, as sparser networks tend to have larger initial MDSs and, consequently, larger difference driver sets, which provide more opportunities for optimization via \CLAPabbr s.

Network topology also plays a critical role. In Erd\H{o}s-R\'enyi (ER-ER) networks \cite{erdos_renyi_1959}, the optimization count declines steadily with increasing density. In contrast, Barab\'asi-Albert (BA-BA) scale-free networks \cite{barabasi_albert_1999} exhibit a distinct peak in optimization potential in the moderately sparse regime ($\langle k \rangle \in [4, 5.9]$), where \algoCLAPS\ saves nearly 70 driver nodes on average. This suggests that the heterogeneous structure of BA networks, with their hubs and varying connectivity, creates a more complex landscape of matching configurations, offering richer opportunities for the coordinated exchanges that \algoCLAPS\ is designed to exploit. The performance on hybrid (ER-BA) networks, as anticipated, falls between that of the pure ER-ER and BA-BA cases.

Beyond its superior optimization effectiveness, \algoCLAPS\ also demonstrates a significant computational advantage. Table~\ref{tab:execution_time_overall} summarizes the average execution times for both algorithms. \algoCLAPS\ is consistently an order of magnitude faster than the \algoRSU\ baseline across all network types. For instance, on ER-ER networks, \algoCLAPS\ requires only about 0.032 seconds on average, compared to \algoRSU's 0.923 seconds. This speed advantage stems from \algoCLAPS's efficient design: it computes the initial matchings only once and then performs a targeted search for \CLAPabbr s. \algoRSU, conversely, is burdened by the high cost of performing $K=20$ randomized matching computations for each layer, followed by $K^2=400$ pairwise comparisons. This combination of superior accuracy and vastly lower computational cost establishes \algoCLAPS\ as a highly efficient and practical algorithm for \UDSabbr\ optimization. The observed variance in execution times, reflected in the standard deviations in Table~\ref{tab:execution_time_overall}, is primarily attributable to the stochastic nature of the graph generation; different network instantiations result in varying sizes of the initial difference driver sets, which in turn affects the number of \CLAPabbr\ searches required to reach the optimal state.

\begin{table*}[!ht]
\centering
\caption{Overall Average Execution Time (seconds $\pm$ Standard Deviation) for Different Algorithms. Times are averaged across all tested average degrees ($\langle k \rangle$) and 10 repetitions for each network type.}
\label{tab:execution_time_overall}
\begin{tabular}{l l c c c c}
\toprule
N & Network Type & RSU (s) & CLAP-S (s) & CLAP-G (s) & ILP (s) \\
\midrule
1000 & BA+BA & 0.796 $\pm$ 0.316 & 0.060 $\pm$ 0.045 & 0.013 $\pm$ 0.005 & 0.737 $\pm$ 0.262 \\
 & ER+BA & 1.137 $\pm$ 0.922 & 0.017 $\pm$ 0.009 & 0.008 $\pm$ 0.005 & 0.820 $\pm$ 0.410 \\
 & ER+ER & 1.383 $\pm$ 1.287 & 0.034 $\pm$ 0.027 & 0.008 $\pm$ 0.005 & 0.844 $\pm$ 0.499 \\
\bottomrule
\end{tabular}%
\end{table*}

\subsubsection{Impact of Network Overlap and Density on \algoCLAPS\ Performance}

\begin{figure*}[!t]
    \centering
    \includegraphics[width=\textwidth]{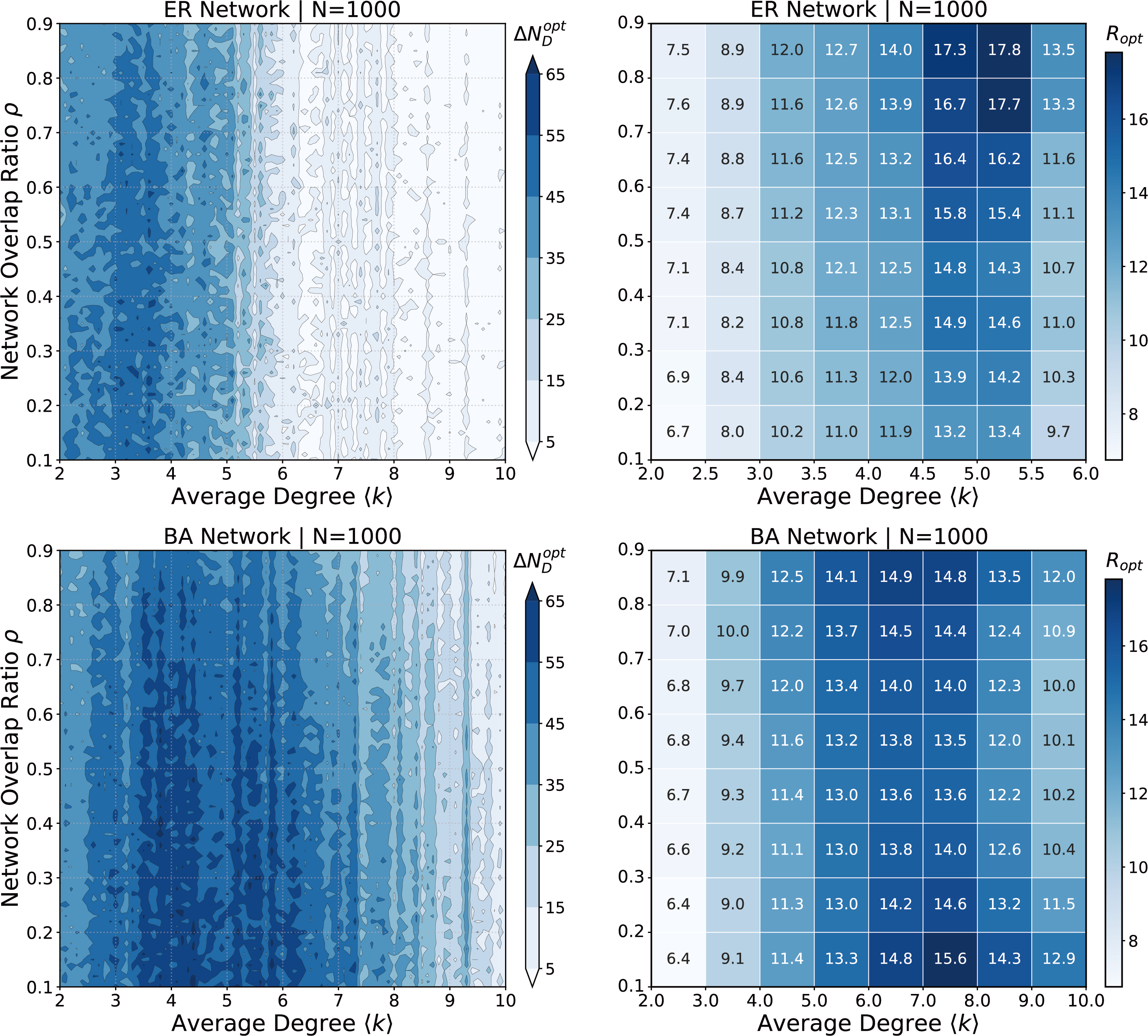}
    \caption{Impact of network overlap ($\rho$) and average degree ($\langle k \rangle$) on \algoCLAPS's performance for ER-ER (top) and BA-BA (bottom) duplex networks ($N=1000$). Left panels: Contour plots of the graph optimization gain, $\Delta N_{D}^{\text{opt}}$. Right panels: Heatmaps of the relative optimization rate, $R_{opt}$ (in \%).}
    \label{fig:overlap_density}
\end{figure*}

To further dissect the performance of our algorithm, we systematically investigated the combined influence of network density (average degree $\langle k \rangle$) and inter-layer structural similarity (network overlap ratio $\rho$). Fig.~\ref{fig:overlap_density} visualizes the graph optimization gain ($\Delta N_{D}^{\text{opt}}$) and the relative optimization rate ($R_{opt}$) across the parameter space for both ER-ER and BA-BA duplex networks.

The analysis reveals that the absolute advantage of \algoCLAPS\ over \algoRSU, $\Delta N_{D}^{\text{opt}}$, is most strongly dictated by network density. For both ER and BA networks, the greatest absolute gains are achieved in sparse to moderately dense regimes, with the advantage diminishing as networks become denser. This is because sparser networks present a larger initial set of difference driver nodes, offering more raw material for optimization. Notably, BA networks sustain a high absolute gain over a broader range of densities compared to ER networks, underscoring \algoCLAPS's effectiveness in exploiting heterogeneous topologies.

A more nuanced picture emerges from the relative optimization rate, $R_{opt}$. In ER networks, $R_{opt}$ exhibits a clear trend of increasing with both higher density (peaking around $\langle k \rangle \in [7, 8]$) and greater overlap. This indicates that while the absolute number of optimizable nodes decreases in these regimes, the remaining optimization problems become disproportionately harder for the \algoRSU\ baseline to solve. \algoCLAPS's guided search proves more robust in identifying the subtle reconfiguration pathways present in denser or more similar layers, thereby maximizing its relative advantage. In BA networks, the relative gain is consistently high across the entire parameter space, often exceeding 12-14\%. This suggests that the inherent structural complexity of scale-free networks creates a persistently challenging landscape for random sampling, allowing \algoCLAPS's deterministic approach to maintain a strong performance edge irrespective of density or overlap. The peak relative performance for BA networks occurs at moderate to high densities, where the interplay of hub-based connectivity and driver node distribution likely creates non-trivial optimization challenges that \algoCLAPS\ is uniquely equipped to resolve.

In summary, these results demonstrate that \algoCLAPS's performance is robust and superior to the baseline across diverse structural conditions. Its absolute gain is primarily driven by network sparsity, while its relative advantage is amplified in scenarios where structural complexity—arising from density, high overlap, or heterogeneous topology—renders random-sampling approaches less effective.

\subsubsection{Relationship between Initial MDS Difference and \algoCLAPS\ Optimization}

\begin{figure}[!ht]
    \centering
    \includegraphics[width=0.92\linewidth]{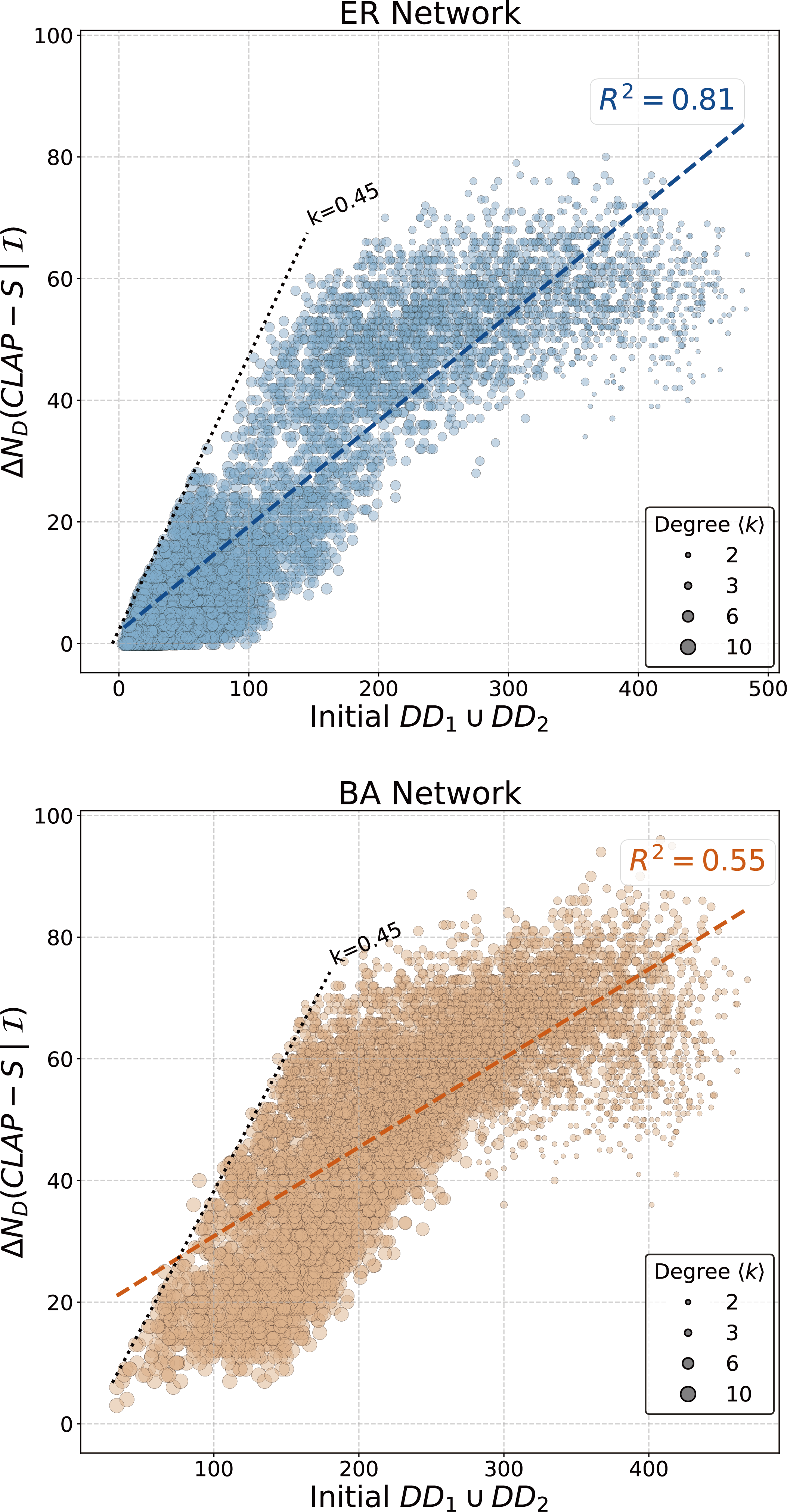}
    \caption{Relationship between the initial difference driver set size ($|\text{DD}_1 \cup \text{DD}_2|$) and the number of nodes optimized by \algoCLAPS\ ($\Delta N_D(\algoCLAPS\ \mid \mathcal{I})$). Each point represents a synthetic duplex network with $N=1000$. Point size indicates average degree $\langle k \rangle$. The dashed lines are linear regression fits for ER networks ($R^2=0.81$) and BA networks ($R^2=0.54$). The dotted line represents a high-efficiency conversion slope of $k=0.45$.}
    \label{fig:scatter_diff_vs_opt}
\end{figure}

To elucidate the primary factors driving \algoCLAPS's performance, we analyze the relationship between the initial disagreement in driver sets and the resulting optimization. Fig.~\ref{fig:scatter_diff_vs_opt} plots the number of nodes saved by \algoCLAPS, $\Delta N_D(\algoCLAPS\ \mid \mathcal{I})$, against the size of the initial difference driver sets, $|\text{DD}_1 \cup \text{DD}_2|$. Each point represents a unique synthetic network configuration, with its size corresponding to the average degree $\langle k \rangle$.

A strong positive correlation is evident for both ER and BA networks, confirming that the initial structural discord between the layers is the main source of optimization potential. The data for ER networks exhibit a remarkably tight linear relationship, with a coefficient of determination $R^2 = 0.81$. This indicates that for homogeneous random graphs, the number of nodes that can be optimized is highly predictable from the initial size of the difference sets. In contrast, the correlation for BA networks, while still strong, is weaker ($R^2 = 0.54$). This greater variance suggests that in scale-free networks, while the initial difference remains the key factor, the complex interplay of hub connectivity and path availability introduces additional factors that influence the final outcome.

The plot also reveals an effective 'conversion rate' of initial difference nodes into optimized nodes. Each successful \CLAPabbr\ execution removes two nodes from the total difference set $|\text{DD}_1 \cup \text{DD}_2|$ and reduces the \UDSabbr\ size by one (if the nodes are distinct) or two (if they become a consistent pair). The dashed black line with slope $k=0.45$ represents a high-efficiency conversion frontier. For ER networks, the linear fit lies close to this frontier, indicating that \algoCLAPS\ efficiently converts a large fraction of the initial difference into \UDSabbr\ reduction. The shallower slope for BA networks suggests a lower average conversion efficiency, likely because the complex paths required for optimization in scale-free structures are more intricate or less frequently available than in random graphs. The network density, represented by point size, primarily determines the initial potential, with sparser networks (smaller points) populating the high-potential, high-optimization region of the plot. Overall, this analysis confirms that \algoCLAPS's performance is fundamentally tied to the initial level of disagreement between the layers, with the underlying network topology modulating the efficiency of the optimization process.

\subsection{Results on Real-World Networks}
\label{subsec:results_real}

To validate the applicability and performance of our algorithm on empirical systems, we applied \algoCLAPS\ to the curated set of 22 real-world duplex networks. We again investigated the relationship between the initial structural disagreement of the layers, measured by $|\text{DD}_1 \cup \text{DD}_2|$, and the resulting optimization achieved by \algoCLAPS, $\Delta N_D(\algoCLAPS\ \mid \mathcal{I})$.

\begin{figure}[!ht]
    \centering
    \includegraphics[width=\linewidth]{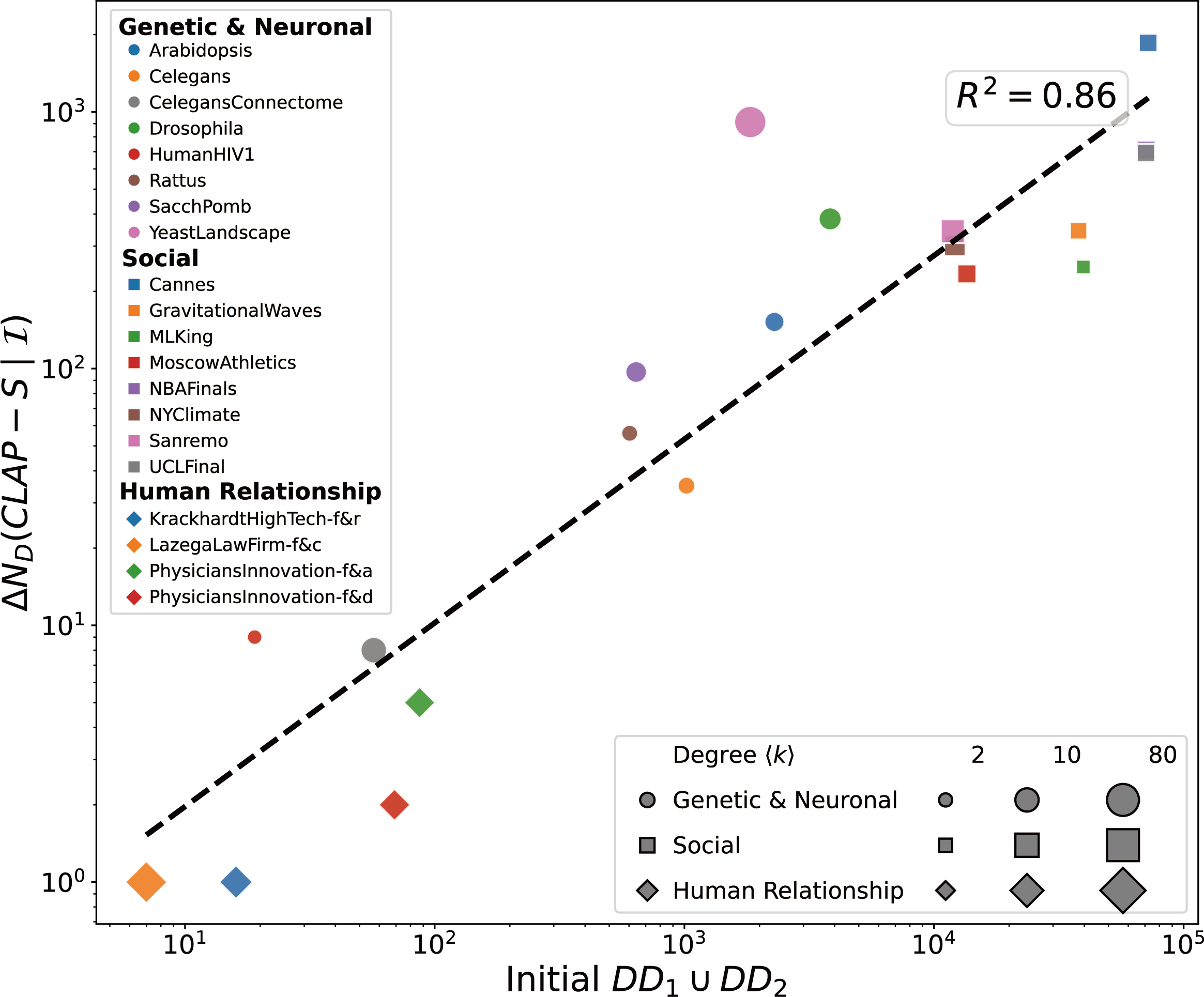}
    \caption{Optimization performance of \algoCLAPS\ on real-world duplex networks. The plot shows the number of nodes optimized by \algoCLAPS, $\Delta N_D(\algoCLAPS\ \mid \mathcal{I})$, as a function of the initial difference driver set size, $|\text{DD}_1 \cup \text{DD}_2|$, on a log-log scale. Each point is a distinct network, categorized by type and sized by its average degree $\langle k \rangle$. The strong linear relationship on the log-log plot ($R^2=0.86$) confirms the predictive power of the initial difference for optimization potential in real-world systems. Note that networks with $|\text{DD}_1|=0$ are not plotted as they offer no optimization space for \algoCLAPS.}
    \label{fig:real_scatter}
\end{figure}

Fig.~\ref{fig:real_scatter} presents this relationship on a log-log scale to accommodate the wide range of network sizes and densities. Each point represents a real-world network, categorized by type and sized by its average degree. The results demonstrate a strong positive correlation between the initial difference in driver sets and the number of nodes \algoCLAPS\ successfully optimizes, achieving a coefficient of determination of $R^2 = 0.86$. This finding powerfully confirms that the principles observed in synthetic networks hold true for real-world systems: a larger initial conflict in layer-specific control requirements provides greater potential for optimization.

the supplementary table “Detailed performance comparison of \algoCLAPS\ and \algoRSU\ on real-world networks”, corroborate these findings. The data consistently show that \algoCLAPS\ finds a smaller \UDSabbr\ ($|\text{\UDSabbr}_{\algoCLAPS}|$) than the \algoRSU\ baseline ($|\text{\UDSabbr}_{\algoRSU}|$) for all networks on which the baseline could be run. For the largest social networks, the \algoRSU\ baseline was computationally prohibitive due to the necessity of repeated maximum matching calculations on massive graphs; these cases are marked with '--' in the table and highlight the critical scalability advantage of \algoCLAPS's targeted optimization. It is also noteworthy that the average \CLAPabbr\ length, $\overline{h}$, is extremely small across all networks where optimization is possible, typically close to 1. This empirically validates the assumption made in our complexity analysis and explains the remarkable efficiency of \algoCLAPS, with execution times ($t_{\algoCLAPS}$) being mere seconds or even milliseconds for all but the largest social networks.

The networks from different domains occupy distinct regions of the plot in Fig.~\ref{fig:real_scatter}, reflecting their intrinsic structural properties. Large-scale biological networks, such as \textit{YeastLandscape} and \textit{Drosophila}, are located in the upper-right quadrant. They possess substantial initial differences and yield correspondingly large optimizations, with \algoCLAPS\ saving hundreds of driver nodes compared to the baseline. For instance, in the \textit{Drosophila} network, \algoCLAPS\ identifies a control set with 352 fewer driver nodes than the baseline could find. This highlights \algoCLAPS's capability to find significant efficiencies in large, complex systems. In a biological context, such a substantial reduction represents a more focused and cost-effective set of potential therapeutic targets, dramatically narrowing the search space for experimental validation in fields like drug discovery or genetic screening. Social networks derived from Twitter, such as \textit{Cannes} and \textit{NBAFinals}, also show considerable optimization potential, falling along the primary trend line and demonstrating reductions of hundreds of nodes. In contrast, smaller human relationship networks, like the \textit{KrackhardtHighTech} and \textit{LazegaLawFirm} datasets, reside in the lower-left. In these cases, the small initial difference sets naturally limit the absolute number of nodes that can be optimized. For instance, in the \textit{KrackhardtHighTech-f\&a} and \textit{LazegaLawFirm-f\&a} networks, the initial difference set $|\text{DD}_1|$ is zero, leaving no space for optimization to begin, which is why they are omitted from Fig.~\ref{fig:real_scatter}. The consistent adherence of these diverse empirical networks to the observed correlation underscores the robustness and practical relevance of the \algoCLAPS\ algorithm for minimizing control costs in real-world multi-layer systems.

\section{Discussion}\label{sec:discussion}

In this work, we addressed the problem of identifying a concise, unified set of driver nodes for duplex networks under the practical engineering constraint of fixed, layer-specific driver budgets. We framed this challenge as a budget-preserving union contraction problem and introduced a novel, path-based framework to solve it. Our main contribution is the \algoCLAPS\ algorithm, a deterministic, polynomial-time procedure that guarantees monotonic improvement and terminates at a provably irreducible state.

\subsection{Summary of Contributions and Positioning}\label{subsec:summary_positioning}

The \algoCLAPS\ algorithm is founded on the \CLAPabbr, a structured operation that reconfigures driver sets across layers without violating pre-defined budgets. Each application of a \CLAPabbr\ strictly reduces the size of the \UDSabbr\ by one, providing a clear potential function for a directed search. The algorithm iteratively finds and applies shortest \CLAPabbr s until no such paths exist. This termination condition, the absence of \CLAPabbr s, serves as a verifiable certificate of a \emph{\CLAPabbr-stable} state.

Our central theoretical result, the \CLAPabbr-Stability Theorem, establishes that such a state is not merely a local fixed point of the algorithm but corresponds to a global minimum of the \UDSabbr\ size over the entire budget-constrained feasible set. This provides a strong guarantee of optimality within a well-defined and practically relevant search space. The experimental results on a wide range of synthetic and real-world networks demonstrate that \algoCLAPS\ is not only effective, achieving significant \UDSabbr\ contraction, but is also highly efficient, consistently outperforming a random-sampling baseline by orders of magnitude in speed.

It is crucial to position our work correctly with respect to related problems in network control and combinatorial optimization.
\textbf{Distinction from General Matching Reconfiguration:} Matching reconfiguration asks whether one can transform one (perfect) matching into another via local moves (e.g., token‑jumping/sliding or 4‑cycle flips), with all intermediate solutions feasible; many variants are PSPACE‑complete or require delicate graph‑class restrictions \cite{ito_reconfiguration_2011, bonamy_perfect_matching_2019, solomon_generalized_matching_reconfiguration_2021}. In contrast, we do \emph{not} reconfigure between two given targets: we search the feasible region of budget‑preserving maximum matchings across layers and apply cross‑layer augmenting paths that \emph{monotonically} decrease the difference mass~$\Delta$, halting with a verifiable certificate of optimality (\CLAPabbr‑stability).
\textbf{Distinction from NP-hard Union Minimization:} Problems involving the minimization of the union of vertex sets, such as the union of a Minimum Driver Set and a Minimum Feedback Vertex Set, are often NP-hard. These problems typically allow for unconstrained trade-offs. Our problem's tractability stems directly from the fixed-budget constraint, which confines the search to the feasible set $\mathcal{M}_1(k_1) \times \mathcal{M}_2(k_2)$ and permits only budget-preserving moves. This difference in problem definition is key to explaining the difference in computational complexity.

\subsection{Implications and Limitations}\label{subsec:implications_limitations}

The concept of a \CLAPabbr-stable state has significant practical implications. It provides an operational criterion to assess whether a given configuration of driver nodes is optimally aligned. For a systems engineer, the absence of a \CLAPabbr\ is a clear signal that no further \UDSabbr\ contraction can be achieved without altering the control budget of at least one layer. This aligns with real-world scenarios where layer-specific performance guarantees (represented by the driver budgets) are non-negotiable.

The primary limitation of our framework is structural rather than algorithmic. The final contracted \UDSabbr\ size is determined by the intrinsic properties of the network layers. While \algoCLAPS\ is guaranteed to find the minimum possible union within the fixed-budget regime, this minimum may not necessarily reach the theoretical lower bound of $\max\{k_1, k_2\}$. Such a bound is only attainable if the network topology permits a perfect nesting of one driver set within another, a condition that is not always met. Our algorithm correctly identifies the true, achievable minimum, whatever it may be.

Another structural limitation is the potential for relay node scarcity. The existence of \CLAPabbr s depends on a sufficient supply of nodes in the CDS and CMS sets to serve as relays. In networks where these sets are small, \CLAPabbr s may be long or non-existent, leading to a state that is \CLAPabbr-stable but may still appear suboptimal from a global, budget-violating perspective.

\subsection{Future Work}\label{subsec:future_work}

This work opens several promising avenues for future research. A natural extension would be to relax the fixed-budget constraint and explore controlled trade-offs. For instance, one could investigate algorithms that allow a small budget increase in one layer if it leads to a disproportionately large decrease in the total \UDSabbr\ size, formulating the problem in a multi-objective optimization framework.

Extending the \CLAPabbr\ framework to networks with more than two layers is another important direction. The core concepts of difference sets and alternating segments could likely be generalized, but the combinatorial complexity of the paths would increase significantly, requiring new algorithmic approaches to maintain tractability. Finally, developing specialized, faster heuristics inspired by the \CLAPabbr\ mechanism could be valuable for real-time applications in extremely large-scale or dynamically evolving networks.

\appendix

\section{Supplementary Materials}

\subsection{Theoretical Framework and Algorithmic Details}\label{app:theory_and_algorithms}

This section provides the detailed theoretical underpinnings and algorithmic components that support the main paper. We begin with a formal review of alternating path properties, followed by the complete proof of the \CLAPabbr-Stability Theorem, and conclude with the pseudocode for the auxiliary procedures used by the \algoCLAPS\ algorithm.

\subsubsection{Properties of Alternating Paths}\label{app:sub_alternating_paths}

We first formalize several key properties of matchings and alternating paths within a single network layer, which are the building blocks of the \CLAPabbr\ framework. Let $\mathcal{B}=(V^+\cup V^-, E_{\mathcal{B}})$ be the bipartite representation of a graph, and let $M$ be a matching in $\mathcal{B}$ \cite{lovasz_matching_1986, schrijver_combinatorial_2003}.

\begin{definition}[Matching and Saturation]\label{def:app_matching}
A \emph{matching} is a set $M \subseteq E_{\mathcal{B}}$ of pairwise vertex-disjoint edges. The saturated vertex set of $M$ is
\[
V_M=\{\,x\in V^{+}\cup V^{-}:\ \exists\, e\in M\text{ with }x\in e\,\}.
\]
A vertex $x$ is \emph{matched} if $x \in V_M$; otherwise, it is \emph{unmatched}.
\end{definition}

\begin{definition}[Alternating Path with Start-Parity]\label{def:app_alt_path}
A simple path $p=(x_0, \dots, x_k)$ in $\mathcal{B}$ is \emph{alternating} with respect to $M$ if its edges alternate between being in $M$ and not in $M$. We say an alternating path starting at $a^- = x_0$ satisfies the \emph{start-parity condition} if its first edge $\{x_0, x_1\}$ belongs to $M$ if and only if $a^-$ is a matched vertex. Formally,
\begin{equation}\label{eq:app_start_parity}
\{x_0,x_1\}\in M\ \Longleftrightarrow\ a^-\in V_M .
\end{equation}
An alternating path satisfying this condition is denoted $p_M(a^-, b^-)$.
See classical references on matching theory for background~\cite{lovasz_matching_1986, schrijver_combinatorial_2003}
\end{definition}

\begin{lemma}[Normalization of Start-Parity]\label{lem:app_start_parity_norm}
Let $M$ be a matching and let $a^-, b^- \in V^-$. If there exists any $M$-alternating path from $a^-$ to $b^-$, then there also exists an $M$-alternating path $p_M(a^-,b^-)$ that satisfies the start-parity condition \eqref{eq:app_start_parity}.
\end{lemma}

\begin{IEEEproof}
If $a^- \notin V_M$, then $a^-$ has no incident edge in $M$. Consequently, any alternating path from $a^-$ must begin with an edge not in $M$, which already satisfies the condition. If $a^- \in V_M$, let $e^\star = \{a^-, y^+\}$ be its unique incident edge in $M$. For any existing alternating path $q$ from $a^-$ to $b^-$, we can construct a new alternating walk by first traversing $e^\star$ to $y^+$ and then following an alternating path from $y^+$ derived from $q$. After removing any induced cycles, we obtain a simple alternating path $p$ from $a^-$ to $b^-$ whose first edge is $e^\star \in M$, thus satisfying the condition.
\end{IEEEproof}

\begin{lemma}[Endpoint Parity]\label{lem:app_endpoint_parity}
Let $p_M(a^-, b^-)$ be an $M$-alternating path that satisfies the start-parity condition. Its last edge belongs to $M$ if and only if the endpoint $b^-$ is a matched vertex ($b^- \in V_M$).
\end{lemma}

\begin{IEEEproof}
Every internal vertex of an alternating path is incident to exactly one edge of $M$ that is on the path. The parity of the path length determines the matching status of the final edge relative to the first. A formal analysis of the path's structure shows that the matching status of the final edge must align with the matching status of the final vertex $b^-$ to avoid violating the matching property at $b^-$ after a symmetric difference operation.
\end{IEEEproof}

\subsubsection{Feasibility of Shortest \CLAPabbr s via Edge-Disjoint Witnesses}
\label{app:shortest_clap_feasible}

\begin{lemma}[Edge-disjointness of witnesses in a shortest \CLAPabbr]
\label{lem:shortest_clap_feasible}
Let $\mathcal{P}=\bigl((v_0 \xrightarrow{\ell_1} v_1),\dots,(v_{k-1} \xrightarrow{\ell_k} v_k)\bigr)$
be a shortest \CLAPabbr. Then, within any fixed layer $\ell\in\{1,2\}$,
the witness $M_\ell$-alternating paths of the segments taken in layer~$\ell$ are pairwise edge-disjoint.
\end{lemma}

\begin{IEEEproof}
Suppose two layer-$\ell$ witness paths share an edge. Let the earlier segment be
$(a \xrightarrow{\ell} b)$ and the later one $(c \xrightarrow{\ell} d)$. Consider the earliest shared edge
on their traces in the $M_\ell$-alternating residual graph. Standard exchange arguments on
alternating paths imply we can shortcut the later segment’s witness to avoid that shared edge
while preserving endpoint parity; iterating removes all shared edges. If a dependency cycle
remained, concatenating the two segments would yield a shorter \CLAPabbr\ (violating minimality).
Hence, witnesses in the same layer are edge-disjoint.
\end{IEEEproof}

\begin{corollary}[Feasibility of shortest \CLAPabbr s]
\label{cor:shortest_clap_feasible}
Every shortest \CLAPabbr\ is feasible in the sense of Definition \CLAPfull\ (in the main text): the layer-wise symmetric differences along its witnesses can be applied without edge conflicts.
\end{corollary}

\subsubsection{Layer-Labeled Meta-Graph and Admissible Orientation}
\label{app:meta_graph_orientation}

\begin{definition}[Layer-labeled meta-graph]
\label{def:app_meta_graph}
Given current state $(M_1,M_2)$ and a comparator $(\widehat M_1,\widehat M_2)$
with $|M_\ell|=|\widehat M_\ell|$, let $H_\ell=(V^+\!\cup V^-, M_\ell\triangle \widehat M_\ell)$.
For each \emph{path} component of $H_\ell$ with $V^-$-endpoints $\{x^-,y^-\}$,
add an undirected edge $\{x,y\}$ to a multigraph $\mathcal K$ and label it by $\ell$.
\end{definition}

\begin{lemma}[Path/cycle structure and label alternation]
\label{lem:app_meta_path_cycle}
Every connected component of $\mathcal K$ is either a simple path or a cycle; along any path,
edge labels alternate $1,2,1,2,\dots$.
\end{lemma}

\begin{lemma}[Unique admissible orientation]
\label{lem:app_admissible_orientation}
For each labeled edge $\{u,v\}$ with label $\ell$, exactly one of
$(u \xrightarrow{\ell} v)$ or $(v \xrightarrow{\ell} u)$ is admissible in the current state.
This defines a unique orientation of $\mathcal K$ into a directed multigraph $\overrightarrow{\mathcal K}$.
\end{lemma}

\begin{IEEEproof}
A $H_\ell$-path alternates between $M_\ell$ and $\widehat M_\ell$; exactly one endpoint has
its $v^-$ unmatched under $M_\ell$. This endpoint is the layer-$\ell$ “driver side” and orients
the edge toward the non-driver endpoint, yielding a unique admissible segment.
\end{IEEEproof}

\subsubsection{Component-wise Impact on Difference Mass and Existence}
\label{app:component_delta_existence}

\begin{lemma}[Endpoint taxonomy and $\Delta$-contribution]
\label{lem:app_component_delta}
Let $\Delta=\Delta(M_1,M_2)$ and $\widehat\Delta=\Delta(\widehat M_1,\widehat M_2)$.
Only path components of $\mathcal K$ whose degree-1 endpoints are of types $(\mathsf L,\mathsf R)$
contribute $+2$ to $\Delta-\widehat\Delta$; all others contribute $\le 0$.
\end{lemma}

\begin{IEEEproof}
Track the XOR of layer-driver bits at endpoints (DD1/DDR, CMS/CDS) and use that internal
nodes flip in both layers (CMS$\leftrightarrow$CDS) contributing zero net change; a short case
enumeration on endpoints yields the stated contributions.
\end{IEEEproof}

\begin{theorem}[Existence of a \CLAPabbr\ when improvement is possible]
\label{thm:app_existence}
If there exists $(\widehat M_1,\widehat M_2)$ with $\widehat\Delta<\Delta$,
then $\overrightarrow{\mathcal K}$ contains a directed label-alternating path from a $\mathrm{DD}_1$
vertex to a $\mathrm{DD}_2$ vertex, i.e., a \CLAPabbr.
\end{theorem}

\begin{IEEEproof}
By Lemma~\ref{lem:app_component_delta}, positive decrease of $\Delta$ implies the presence
of a path component with $(\mathsf L,\mathsf R)$ endpoints; orienting edges by
Lemma~\ref{lem:app_admissible_orientation} yields a directed alternating path—a \CLAPabbr.
\end{IEEEproof}

\subsubsection{Proof of the \CLAPabbr-Stability Theorem}\label{app:sub_proof_stability}

We present the complete proof for Theorem \CLAPabbr-or-Optimal (in the main text), which establishes that a state $(M_1, M_2)$ is \CLAPabbr-stable if and only if it minimizes the \UDSabbr\ size $|U(M_1, M_2)|$ over the feasible set $\mathcal{M}_1(k_1) \times \mathcal{M}_2(k_2)$.

The proof proceeds by contradiction. We assume a state $(M_1, M_2)$ is \CLAPabbr-stable but is \emph{not} a minimum. This implies the existence of another state $(\widehat{M}_1, \widehat{M}_2)$ in the feasible set with a strictly smaller \UDSabbr\ size. We then show that the existence of such a "better" state necessitates the existence of at least one \CLAPabbr\ in the original state, which contradicts our initial assumption of \CLAPabbr-stability. To establish this, we construct a layer-labeled meta-graph based on the symmetric differences of the matchings.

\paragraph{Layer-wise Symmetric Differences and Node Signatures}
Let $(M_1, M_2)$ be the current state and $(\widehat{M}_1, \widehat{M}_2)$ be any other state in the feasible set. Let $D_\ell = D_\ell(M_\ell)$ and $\widehat{D}_\ell = D_\ell(\widehat{M}_\ell)$. For each layer $\ell \in \{1,2\}$, we define the symmetric-difference graph in the bipartite representation as $H_\ell = (V^+\cup V^-, M_\ell \triangle \widehat{M}_\ell)$. Each connected component of $H_\ell$ is either an even cycle or a simple path whose edges alternate between $M_\ell$ and $\widehat{M}_\ell$.

To precisely classify how each node's driver status changes between the two states, we define a signature $\delta(v) = (\delta_1(v), \delta_2(v)) \in \{-1, 0, +1\}^2$, where
\[
\delta_\ell(v) = \mathbf{1}_{\{v \in \widehat{D}_\ell\}} - \mathbf{1}_{\{v \in D_\ell\}}.
\]
This signature provides a complete taxonomy of node-state changes from the perspective of the current state $(M_1, M_2)$:
\[
\begin{alignedat}{2}
\mathsf L &= (-1,0) &\quad& \text{loses layer 1 driver ($v\in D_1$)},\\[1pt]
\mathsf R &= (0,-1) &\quad& \text{loses layer 2 driver ($v\in D_2$)},\\[1pt]
\mathsf N_1 &= (+1,0) &\quad& \text{gains layer 1 driver ($v\notin D_1$)},\\[1pt]
\mathsf N_2 &= (0,+1) &\quad& \text{gains layer 2 driver ($v\notin D_2$)},\\[1pt]
\mathsf C^+ &= (+1,+1) &\quad& \text{CMS}\to\text{CDS (consistent)},\\[1pt]
\mathsf C^- &= (-1,-1) &\quad& \text{CDS}\to\text{CMS (consistent)},\\[1pt]
\mathsf X^+ &= (-1,+1) &\quad& \text{DD}_1\to\text{DD}_2\text{ (difference-side flip)},\\[1pt]
\mathsf X^- &= (+1,-1) &\quad& \text{DD}_2\to\text{DD}_1\text{ (difference-side flip)},\\[1pt]
\mathsf Z &= (0,0) &\quad& \text{unchanged}.
\end{alignedat}
\]

\paragraph{The Layer-Labeled Meta-Graph}
We now construct an undirected multigraph $\mathcal{K}=(V, \mathcal{E})$ that captures the structural relationship between the states $(M_1, M_2)$ and $(\widehat{M}_1, \widehat{M}_2)$.

\begin{definition}[Layer-Labeled Meta-Graph]\label{def:meta_graph}
For each layer $\ell \in \{1,2\}$ and for each \emph{path} component of the symmetric difference graph $H_\ell$ with endpoints $x^-, y^-$, we add an undirected edge $\{x, y\}$ to the edge set $\mathcal{E}$ and assign it the layer label $\lambda(\{x,y\}) = \ell$. Cycle components of $H_\ell$ do not contribute edges to $\mathcal{K}$.
\end{definition}

\begin{lemma}[Meta-Graph Structure]\label{lem:meta_structure}
Every connected component of the meta-graph $\mathcal{K}$ is either a simple path or a simple cycle. Furthermore, along any path component, the edge labels must alternate (e.g., $1, 2, 1, 2, \dots$).
\end{lemma}

\begin{IEEEproof}
By construction, an edge is added to $\mathcal{K}$ for each path component in each $H_\ell$. A node $v \in V$ can be an endpoint of at most one path component in $H_1$ and at most one in $H_2$. Therefore, the degree of any node $v$ in $\mathcal{K}$, $\deg_{\mathcal{K}}(v)$, is at most 2. A graph where all node degrees are at most 2 must be a disjoint union of simple paths and simple cycles. The alternating label property follows because two consecutive edges in a path in $\mathcal{K}$ cannot have the same label. If they did, their shared central node would have a degree of 2 contributed by a single layer, which is impossible by the construction, as a node can be an endpoint of only one path component per layer.
\end{IEEEproof}

\paragraph{From Meta-Graph Paths to \CLAPabbr s}
Each edge $\{u,v\}$ in $\mathcal{K}$ with label $\ell$ corresponds to an $M_\ell$-alternating path between $u^-$ and $v^-$ in the bipartite graph $\mathcal{B}_\ell$. Exactly one of these endpoints, say $u$, must be in $D_\ell(M_\ell)$, while the other, $v$, is not. This configuration defines a unique \emph{admissible orientation} for the edge, from the driver to the non-driver, which corresponds to an admissible segment $(u \xrightarrow{\ell} v)$. By orienting all edges in $\mathcal{K}$ in this manner, we obtain a directed graph $\overrightarrow{\mathcal{K}}$. A directed, label-alternating path in $\overrightarrow{\mathcal{K}}$ from a node in $\text{DD}_1$ to a node in $\text{DD}_2$ directly translates to a valid \CLAPabbr.

\paragraph{Existence of Improving \CLAPabbr s}
Let $\Delta = \Delta(M_1, M_2)$ and $\widehat{\Delta} = \Delta(\widehat{M}_1, \widehat{M}_2)$. The change in the difference mass, $\Delta_{change} = \Delta - \widehat{\Delta}$, can be analyzed component-wise in the meta-graph $\mathcal{K}$.

\begin{lemma}[Component-wise Contribution to $\Delta$ Change]\label{lem:component_contribution}
\begin{enumerate}\itemsep 1pt
    \item Any cycle component in $\mathcal{K}$ contributes $0$ to $\Delta_{change}$.
    \item Any path component in $\mathcal{K}$ whose endpoints are of types $(\mathsf{L}, \mathsf{R})$ contributes exactly $2$ to $\Delta_{change}$.
    \item Any path component whose endpoints involve at least one "neutral gain" type, such as $(\mathsf{L}, \mathsf{N}_2)$ or $(\mathsf{N}_1, \mathsf{N}_2)$, contributes $0$ to $\Delta_{change}$.
    \item All other path components contribute non-positive values to $\Delta_{change}$.
\end{enumerate}
\end{lemma}

\begin{IEEEproof}
The change in difference mass is given by the sum over all nodes of the change in the value of $(\mathbf{1}_{v \in D_1} \oplus \mathbf{1}_{v \in D_2})$, where $\oplus$ is the exclusive OR. Internal nodes of a path in $\mathcal{K}$ have degree 2, meaning their driver status flips in \emph{both} layers (e.g., they transition between CMS and CDS). This flip does not change their contribution to the difference mass, since $(0 \oplus 0) = 0$ and $(1 \oplus 1) = 0$. Therefore, the net change comes only from the degree-1 endpoints of the path components. A direct case analysis of the endpoint types confirms the contributions:
\begin{itemize}
    \item An $\mathsf{L}$ endpoint is a DD1 node ($1\oplus0=1$) that becomes a CMS node ($0\oplus0=0$). The change is $0-1 = -1$. Its contribution to $\Delta_{change}$ is $-(-1) = +1$.
    \item An $\mathsf{R}$ endpoint is a DD2 node ($0\oplus1=1$) that becomes a CMS node ($0\oplus0=0$). The change is $0-1 = -1$. Its contribution to $\Delta_{change}$ is $+1$.
    \item An $\mathsf{N}_1$ endpoint is a CMS node ($0\oplus0=0$) that becomes a DD1 node ($1\oplus0=1$). The change is $1-0 = +1$. Its contribution to $\Delta_{change}$ is $-1$.
    \item An $\mathsf{N}_2$ endpoint is a CMS node ($0\oplus0=0$) that becomes a DD2 node ($0\oplus1=1$). The change is $1-0 = +1$. Its contribution to $\Delta_{change}$ is $-1$.
\end{itemize}
A path with $(\mathsf{L}, \mathsf{R})$ endpoints thus contributes $1+1=2$ to $\Delta_{change}$. A path with $(\mathsf{L}, \mathsf{N}_2)$ endpoints contributes $1-1=0$. All other combinations yield a non-positive sum.
\end{IEEEproof}

\begin{IEEEproof}[Proof of Theorem \CLAPabbr-or-Optimal]
We prove that a state is \CLAPabbr-stable if and only if it is a minimum.
\begin{itemize}
    \item[($\Rightarrow$)] We prove the contrapositive: if a state $(M_1, M_2)$ is not a minimum, then a \CLAPabbr\ must exist. If $(M_1, M_2)$ is not a minimum, there exists another feasible state $(\widehat{M}_1, \widehat{M}_2)$ such that $|U(\widehat{M}_1, \widehat{M}_2)| < |U(M_1, M_2)|$. By Proposition Equivalence of Objectives (in the main text), this implies $\widehat{\Delta} < \Delta$, so $\Delta_{change} = \Delta - \widehat{\Delta} > 0$. By Lemma \ref{lem:component_contribution}, a positive total change requires the existence of at least one path component in the meta-graph $\mathcal{K}$ that contributes positively. This must be a path with $(\mathsf{L}, \mathsf{R})$ endpoints. An $\mathsf{L}$ endpoint corresponds to a starting node in $\text{DD}_1$, and an $\mathsf{R}$ endpoint corresponds to a terminal node in $\text{DD}_2$. As established, such a path in the meta-graph directly translates to a valid \CLAPabbr\ from $\text{DD}_1$ to $\text{DD}_2$ in the state $(M_1, M_2)$. Therefore, the state is not \CLAPabbr-stable.
    \item[($\Leftarrow$)] We prove that if a state is not \CLAPabbr-stable, it is not a minimum. If a state is not \CLAPabbr-stable, then by definition, at least one \CLAPabbr\ exists. By the \CLAPabbr\ Gain Theorem (in the main text), applying this \CLAPabbr\ produces a new feasible state $(M_1', M_2')$ with $|U(M_1', M_2')| = |U(M_1, M_2)| - 1$. Since a state with a strictly smaller \UDSabbr\ size exists, the original state $(M_1, M_2)$ cannot be a minimum.
\end{itemize}
Combining both directions, we conclude that a state is \CLAPabbr-stable if and only if it is a minimum within the feasible set $\mathcal{M}_1(k_1) \times \mathcal{M}_2(k_2)$.
\end{IEEEproof}

\subsubsection{Auxiliary Algorithmic Procedures}\label{app:sub_aux_algorithms}

The main \algoCLAPS\ algorithm relies on several subroutines. Their pseudocode is provided in Algorithm \ref{alg:app_aux_procedures}. These functions handle the low-level graph traversal and matching manipulations required to find and apply \CLAPabbr s.

\begin{algorithmic}[1]
\label{alg:app_aux_procedures}
\Statex \textbf{Procedure } \textsc{AltReach}$({\cal S} \mid M_\ell)$
\Statex \quad \emph{Purpose.} Given a \emph{set of sources} ${\cal S}\subseteq V$ and matching $M_\ell$ in layer $\ell$, return the full set
\begin{equation*}
\begin{aligned}
R = \{\, v\in V \setminus \mathcal{S} \;:\;& \exists \text{ an } \ell\text{-alternating path}\\
&\text{from some } u\in\mathcal{S} \text{ to } v \,\}.
\end{aligned}
\end{equation*}
\Statex \quad \emph{Implementation.} Run a single \emph{multi-source} BFS in the $M_\ell$-alternating residual graph, seeding all $u\in{\cal S}$ simultaneously, and respecting the start-parity for layer $\ell$. Record parent pointers for each reached $v$ to enable witness reconstruction.
\State \Return $R$

\Statex

\Statex \textbf{Procedure } \textsc{AltPath}$(u, v \mid M_\ell)$
\Statex \quad \emph{Purpose.} Return a \emph{shortest} $M_\ell$-alternating path $p$ from $u$ to $v$ under the layer-$\ell$ move pattern, or $\bot$ if no such path exists. One may realize this with a BFS that records parent pointers; details are omitted here.
\State \Return $p$ or $\bot$

\Statex

\Statex \textbf{Procedure } \textsc{UnrollSegments}$((u,\ell), v, \mathrm{pred},\mathrm{predLayer})$
\State $\mathit{seg}\gets[\,]$;\; $(x,\eta)\gets (u,\ell)$
\While{$\mathrm{pred}(x,\eta)\neq\text{nil}$}
  \State $(p,\pi)\gets \mathrm{pred}(x,\eta)$ 
  \State append $(p, x, \pi)$ to $\mathit{seg}$;\; $(x,\eta)\gets (p,\pi)$
\EndWhile
\State $\mathit{seg} \gets \text{reverse}(\mathit{seg})$
\State \textit{// append the final segment and build witnesses for all segments}
\State append $(u, v, \ell)$ to $\mathit{seg}$
\State $\mathit{segmentsWithWitnesses}\gets [\,]$
\For{$(a,b,\lambda)$ in $\mathit{seg}$}
  \State $p \gets \textsc{AltPath}(a,b \mid M_\lambda)$ \Comment{shortest alternating path with correct endpoint-parity}
  \State append $(a,b,\lambda,\ E(p))$ to $\mathit{segmentsWithWitnesses}$
\EndFor
\State \Return $\mathit{segmentsWithWitnesses}$

\Statex

\Statex \textbf{Procedure } \textsc{SymDiff}$(M_\ell, E(p))$
\State \Return $(M_\ell \setminus E(p)) \cup (E(p) \setminus M_\ell)$

\Statex

\Statex \textbf{Procedure } \textsc{ApplyCLAP}$(\mathit{segments}, M_1,M_2,D_1,D_2)$
\For{$(x,y,\ell,E(p))$ in $\mathit{segments}$ in order}
  \If{$\ell=1$}
     \State $M_1 \gets (M_1 \setminus E(p)) \cup (E(p)\setminus M_1)$
     \State $D_1 \gets (D_1\setminus\{x\})\cup\{y\}$
  \Else
     \State $M_2 \gets (M_2 \setminus E(p)) \cup (E(p)\setminus M_2)$
     \State $D_2 \gets (D_2\setminus\{y\})\cup\{x\}$
  \EndIf
\EndFor
\State \Return $(M_1,M_2,D_1,D_2)$

\end{algorithmic}

\subsection{Detailed Experimental Results}\label{app:detailed_results}

This section contains the full experimental data for the real-world duplex networks analyzed in the main paper. Table \ref{tab:app_real_world_results} provides a comprehensive breakdown of the performance of the \algoCLAPS\ algorithm and the \algoRSU\ baseline, supporting the analysis presented in Section Results on Real-World Networks (in the main text).

\begin{table*}[!htbp]
\caption{
    Detailed performance comparison of \algoCLAPS\ and other algorithms on real-world networks. 
    $|U|_0$ is the initial \UDSabbr\ size. 
    $|U|_{\text{CLAPS}}$, $|U|_{\text{RSU}}$, $|U|_{\text{CLAPG}}$, and $|U|_{\text{ILP}}$ are the final sizes after running the respective algorithms. 
    $\overline{h}$ is the average \CLAPabbr\ length found by \algoCLAPS.
    $t$ is the execution time in seconds. 
    Execution times exceeding 500 seconds are marked with '--'. 
    ILP was not run for networks with more than 10000 nodes, also marked with '--'.
    Cases where the algorithm could not run or the result was invalid are marked with '--'.
}
\label{tab:app_real_world_results}
\centering
\begin{tabular}{lrrrrrrrrrr}
\toprule
Network Name & $|U|_0$ & $|U|_{\text{CLAP-S}}$ & $|U|_{\text{RSU}}$ & $|U|_{\text{CLAP-G}}$ & $|U|_{\text{ILP}}$ & $\overline{h}$ & $t_{\text{CLAP-S}}$ & $t_{\text{RSU}}$ & $t_{\text{CLAP-G}}$ & $t_{\text{ILP}}$ \\
\midrule
Arabidopsis & 6593 & 6453 & 6580 & 6467 & 6453 & 1.164 & 0.356 & 2.818 & 0.097 & 5.360 \\
Celegans & 3143 & 3114 & 3138 & 3117 & 3114 & 1.103 & 0.017 & 0.939 & 0.010 & 0.985 \\
Drosophila & 7383 & 7000 & 7342 & 7029 & 7000 & 1.149 & 1.850 & 3.652 & 0.401 & 11.642 \\
HumanHIV1 & 986 & 977 & 983 & 978 & 977 & 1.111 & 0.001 & 0.154 & 0.001 & 0.294 \\
SacchPomb & 2366 & 2267 & 2350 & 2276 & 2267 & 1.263 & 0.266 & 0.951 & 0.037 & 1.394 \\
Rattus & 2443 & 2381 & 2438 & 2389 & 2381 & 1.177 & 0.105 & 0.791 & 0.024 & 1.006 \\
CelegansConnectome & 61 & 55 & 59 & 58 & 55 & 1.833 & 0.009 & 0.238 & 0.000 & 0.174 \\
YeastLandscape & 3659 & 2757 & 3620 & 2757 & 2757 & 1.018 & 3.146 & 2.790 & 3.257 & 36.507 \\
Cannes & 421123 & - & 421079 & 419301 & - & - & - & 263.499 & 160.982 & - \\
MLKing & 321966 & 321726 & 321945 & 321731 & - & 1.021 & 232.491 & 150.798 & 12.282 & - \\
MoscowAthletics & 86384 & 86162 & 86359 & 86167 & - & 1.041 & 31.315 & 33.352 & 3.075 & - \\
NYClimate & 99437 & 99154 & 99418 & 99163 & - & 1.032 & 42.636 & 44.494 & 4.197 & - \\
NBAFinals & 741850 & - & 741810 & 741140 & - & - & - & 447.346 & 107.869 & - \\
Sanremo & 53817 & 53491 & 53802 & 53500 & - & 1.034 & 9.916 & 30.789 & 1.955 & - \\
UCLFinal & 672094 & - & 672072 & 671460 & - & - & - & 384.775 & 91.643 & - \\
GravitationalWaves & 357139 & 356772 & 357102 & 356779 & - & 1.022 & 339.581 & 172.306 & 23.684 & - \\
KrackhardtHighTech-f\&a & 2 & 2 & 2 & 2 & 2 & 0 & 0.000 & 0.025 & 0.000 & 0.011 \\
KrackhardtHighTech-f\&r & 16 & 16 & 16 & 16 & 16 & 0 & 0.000 & 0.015 & 0.000 & 0.007 \\
LazegaLawFirm-f\&a & 6 & 6 & 6 & 6 & 6 & 0 & 0.000 & 0.103 & 0.000 & 0.046 \\
LazegaLawFirm-f\&c & 7 & 6 & 6 & 6 & 6 & 1 & 0.000 & 0.053 & 0.000 & 0.053 \\
PhysiciansInnovation-f\&a & 127 & 120 & 122 & 120 & 120 & 1 & 0.002 & 0.062 & 0.000 & 0.050 \\
PhysiciansInnovation-f\&d & 103 & 99 & 100 & 99 & 99 & 1 & 0.001 & 0.192 & 0.000 & 0.051 \\
\bottomrule
\end{tabular}
\end{table*}

\noindent From Table~\ref{tab:app_real_world_results} we make three observations.

(i) \emph{Quality vs.\ ILP (normal-scale graphs).} For every dataset on which the exact ILP solver was executed (i.e., up to $\sim$10K nodes in our setting), the union size returned by \algoCLAPS\ exactly matches the ILP optimum (e.g., \textit{Arabidopsis}, \textit{Celegans}, \textit{Drosophila}, \textit{YeastLandscape}, and all small social networks), confirming the optimality certificate discussed in the main text (Section “Results on Real-World Networks”). Compared to \algoRSU, the gap can be substantial (e.g., \textit{YeastLandscape}: $3620\!\to\!2757$; \textit{Drosophila}: $7342\!\to\!7000$).

(ii) \emph{Runtime.} \algoCLAPS\ typically finishes in seconds and is often faster than \algoRSU on these graphs (e.g., \textit{Drosophila}: 1.85\,s vs.\ 3.65\,s; \textit{SacchPomb}: 0.27\,s vs.\ 0.95\,s), while remaining comparable in a few cases where \algoRSU benefits from its sampling budget (e.g., \textit{YeastLandscape}: 3.15\,s vs.\ 2.79\,s). The average \CLAPabbr\ length $\overline{h}$ stays close to $1$ (typically $1.02$–$1.26$), explaining the good scaling behavior; outliers such as \textit{CelegansConnectome} (1.83) remain small in absolute terms.

(iii) \emph{Massive graphs ($\ge\!3\times 10^5$ nodes).} By design we do not run ILP on these instances; \algoCLAPS\ may exceed the 500\,s cap (e.g., \textit{Cannes}, \textit{NBAFinals}, \textit{UCLFinal}). In such cases, the greedy variant \algoCLAPG\ is a practical substitute that still yields strong unions while being significantly faster than \algoRSU: \textit{Cannes} ($419{,}301$ vs.\ $421{,}079$ in 161\,s vs.\ 263\,s), \textit{NBAFinals} ($741{,}140$ vs.\ $741{,}810$ in 108\,s vs.\ 447\,s), and \textit{UCLFinal} ($671{,}460$ vs.\ $672{,}072$ in 91.6\,s vs.\ 384.8\,s). Overall, on normal-scale networks \algoCLAPS\ achieves ILP-optimal unions with much lower runtime; on massive networks, \algoCLAPG\ preserves most of the quality gains at a fraction of the time.

\ifCLASSOPTIONcompsoc
  \section*{Acknowledgments}
\else
  \section*{Acknowledgment}
\fi
This work was supported in part by the National Natural Science Foundation of China (NSFC) under Grant~62176129; 
the NSFC--Jiangsu Joint Fund under Grant~U24A20701; 
and the Hong Kong RGC Strategic Target Grant (Grant No.~\mbox{STG1/M-501/23-N}).

\bibliographystyle{IEEEtran}
\bibliography{ref}

\end{document}